\newcommand{\be}{\begin{equation}}
\newcommand{\ee}{\end{equation}}
\newcommand{\ef}{\end{figure}}
\newcommand{\f}{\begin{figure}}
\newcommand{\qea}{q_{EA}}
\newcommand{\qdo}{q_{13}}
\newcommand{\qve}{q_{23}}
\newcommand{\qtr}{q_{12}}
\newcommand{\qb}{q^{fix}}
\newcommand{\ddo}{d_{13}}
\newcommand{\dve}{d_{23}}
\newcommand{\dtr}{d_{12}}

\newcommand{\nin}{\noindent}

\documentstyle[12pt]{article}
\input{epsfig.sty}
\begin{document}

\title {
  Study of the ultrametric structure
 of Finite Dimensional Spin Glasses throught a constrained Monte Carlo dynamic}

\author{A. Cacciuto\\
{\small Dipartimento di Fisica and Infn, Universit\`a di Cagliari}\\
 {\small Via Ospedale 72, 09100 Cagliari (Italy) }\\
  {\small E-mail   cacciuto@vaxca.ca.infn.it}}

\maketitle

\begin{abstract}
 We report a very detailed relation about the study, by a 
constrained Monte Carlo
 dynamic  of a 4D EA spin glass model ($J=\pm 1$). In particular we
 concentrate our attention on the study of the behaviour 
 of the system under different dynamical parameters 
 in order to optimize our analysis and  try to understand
 the low energy states structure. We find that in the thermodynamic limit 
 this structure assumes just  ultrametric features as already 
 established for the SK model in the SRSB (Spontaneous Replica
 Simmetry Breaking) theory.
\end{abstract}

\section*{\bf Introduction}
At present there are two different and contrasting approaches about 
the description of the spin-glass phase of a finite-dimensional EA model.
The first, the replica approach, follows the solutions of the SK model
describing the nature of spin-glass phase at finite dimensions
 similar  to the one of the  mean-field theory \cite{beyond}. 
The second approach is based on the phenomenological theory of droplets
 and predicts a spin-glass phase dominated by only one equilibrium state 
\cite{fisher}.
The difficulties in the research of analytic results in this field
imposes that a large part of the work is delegated 
to the numerical simulations.
A large number of works (\cite{mari}\cite{sim}\cite{4d}\cite{Young})
 supports the 
replica approach which
 at present seems the more appropriate for the description
of realistic spin-glasses.

In this work we study the four-dimensional Ising Spin Glass at zero external
magnetic field by means of Monte Carlo numerical simulations for several
small sizes $N=L^4$ with L=3, 4, 5, 6, 7, 8 with coupling quenced $\pm J$.

In detail we will concentrate our attention on the structure of low 
energy states of the model by a constrained dynamic that will be
here examinated   very in detail  (see also \cite{nostro}).
In the first and second section we will briefly mention  the analytic results
obtained with the SK model and we will add some comments about the  main
 numerical works undertaken about the ultrametric problem in spin-glasses
(see also \cite{micro1}).
In section 3, we will introduce the metric used for this model.
Sections 4, 5, and 6 are dedicated to 
the description of the simulated system
and to the Monte Carlo constrained dynamic realized.
Finally the last sections report the results of the  numerous simulations
effected.  
\section{The ultrametricity in the mean field theory}
Rigorous analytic studies (\cite{beyond}\cite{Parisi2}\cite{micro})
 have  clearly demonstrated 
the ultrametric structure of low energy states of the SK model.
The starting point of this analysis is the calculation 
of the probability for  three pure states $\alpha_{1},\alpha_{2},\alpha_{3} $
to have mutual overlaps
\be
q_{1}=q^{\alpha_{2}\alpha_{3}},\hspace{.1cm}  q_{2}=q^{\alpha_{3}\alpha_{1}},
\hspace{.1cm}
q_{3}=q^{\alpha_{1}\alpha_{2}}\ .
\ee
One can demonstrate \cite{Parisi2} that subsists the relation

\vspace{.5cm}
$P(q_{1},q_{2},q_{3})=\frac{1}{2}P(q_{1})x(q_{1})\delta (q_{1}-q_{2})
\delta (q_{1}-q_{3})+$
\begin{equation}
+\frac{1}{2}\{P(q_{1})P(q_{2})
\theta (q_{1}-q_{2})\delta(q_{2}-q_{3})+\mbox{2\hspace{.1cm}permutations}\ \}\ .
\label{ultrametr}
\end{equation}
The study of this equation \cite{Parisi2}
tells us that $P(q_{1},q_{2},q_{3})$
is  null except when two of the overlaps are equals and not larger of
the third one.
If we define the distance between two pure states as
\begin{equation}
d^{2}_{\alpha\beta}=\frac{1}{N} \sum_{i} (m_{i}^{\alpha}-m_{i}^{\beta})^{2}\ ,
\end{equation}
\noindent
(this is not the only way), can be easily verified that
\begin{equation}
d^{2}_{\alpha\beta}=2(q_{EA}-q_{\alpha\beta})\ .
\end{equation}
\noindent
In this contest the equation (\ref{ultrametr}) establishes that 
triangles builded with three configurations, chosen in agreement with their
Boltzmann weights, are always either
 equilaters or isosceles, and in this last case 
 the different side is the smallest. In this prospective, we need to
 replace the triangular   inequality
\begin{equation}
d_{ab}\leq d_{ac}+d_{bc}
\end{equation}
\noindent
with the more restrictive condition 
\begin{equation}
d_{ab}\leq max(d_{ac},d_{bc})\ .
\end{equation}
\noindent

A space with this metric is called  {\bf ultrametric}.

\section{\bf Ultramericity in literature} 

Attempts of analysis with calculator of ultrametric properties of finite
dimensional  spin  glasses are very few in literature (\cite{u1}\cite{4d})
and have not given any decisive contribute. The more 
remarkable results of interest for our work are 
probably those obtained by Ciria, Parisi and Ritort \cite{4d} in 4 dimensions
at T=1.4. Now we discuss the crucial elements of the philosophy followed
in that work.
For every triad of three overlaps that one can build (see also section 4)
from three different replicas of the system (with the same disorder
realization but an autonomous dynamic) one analyses, for a fixed
 bigger overlap in an established range of values,
 the difference between the middle overlap and
the smallest excluding terms where the triangular inequality is violated
from finite size corrections.
For this purpose one defines for every overlap a distance
\be
d_{\alpha\beta}=[2(q_{max}-q^{\alpha\beta})]^{1/2}\ ,
\ee
\noindent
where $q_{max}$ is the largest overlap relative to the thermodynamic limit
\be
q_{max}=\lim_{L\rightarrow\infty}q^{L}_{max}\ ,
\ee
(which can be obtained by the scaling relation
$P(q>q_{max})\sim P_{max}\cal F(N(q-q_{max})^{\frac{D}{dq}}) \cite{4d})$, and
one consider only the overlap triads that satisfy the inequality
\be
d_{max}\leq d_{min}+d_{middle}\ .
\ee
\noindent
The results of this analysis have not given a decisive answer to the problem.
 In effect if we observe figure (\ref{u4d}), where is indicated
in abscissas the value of  $\delta q=q_{medio}-q_{min}$ and in ordinates
 $P(\delta q)$, we note that a dependence on the curves from the volume
 is not determined with sufficient
statistic precision. 
By these simulation people thought that it were possible
to  increase the volume size
to obtain a precise indication of a ultrametric behaviour 
 in the infinite volume limit.
Similar works about the SK model \cite{usk1}\cite{usk2} have given
very satisfying results pointing out clearly the ultrametric
structure of the states. The first of these works is from Bhatt 
Young \cite{usk1}. The result is shown in figure  (\ref{pqmid12}). 
In this case the curve tends explicitly to narrow increasing the lattice 
volume showing the existence of the predicted structure between the states.

\section{\bf The distance}
There are  two big problems for a direct analysis of the ultrametric structure:

1) There are corrections for small N that are not known analytically
 
2) The triangular inequality can impose some bonds that are not easily 
   distinguishable from the more restrictive ones provided from
   the ultrametric inequality.

To understand the last affirmation we take a triangle with sides   
$d_1$, $d_2$, $d_3$ and let   $d_3$ the smallest one. 
The triangular inequality imposes that 
\begin{equation}
|d_2-d_3|\leq d_1\leq d_2+d_3
\end{equation}
\noindent
to compare with the ultrametric request
\begin{equation}
d_1=d_2\ .
\end{equation}
\noindent
A correction due to the limited value of N causes some violations in
 the last equation
\begin{equation}
d_1=d_2\pm O(1/N^\delta)\ .
\end{equation}
\noindent
If $d_3$ is fairly little, the imposed conditions  from the two inequality
are practically indistinguishable. To get over these problems it is necessary
to define a metric which makes the two conditions as distinct as possible.

We define (see also \cite{nostro})
the square distance between the two pure states with overlap $q$ as
\begin{equation}
d^2=\frac{q_{EA}-q}{2q_{EA}}\ ,
\end{equation}
\noindent
so when $q=-q_{EA}$ we will have $d^2=1$, the lowest overlap imposes
a largest square distance, while for $q=q_{EA}$ we have that $d^2=0$,
the largest overlap involves a lowest square distance.
The triangular inequality imposes that when we take three pure states   
the mutual distances satisfy the relation 
\be
  d_{13} + d_{23} \ge d_{12}\ .
\ee
Squaring both members we have
\be
  (d_{13}+d_{23})^2 \ge d_{12}^2\
\label{triangolare}
\ee
\noindent
and so

\vspace{.5cm}
\be
 \frac{(\qea-\qdo)}{2\qea}
 + \frac{(\qea-\qve)}{2\qea}+
 2 \frac{\sqrt{\qea-\qdo}\sqrt{\qea-\qve}}{2\qea}
 \ge \frac{\qea-\qtr}{2\qea}\ .
\ee
We fix now two of the three overlaps, for exemple
\be
  \qdo = \qve \equiv \qb\
\ee
\noindent
(we take for convenience of notation $d_{12}\equiv d$ and 
$q_{12}\equiv q$).\\
From the equation (\ref{triangolare}) we have
\be
  4(\qea - \qb) \ge \qea - q \ge 0\ ,
\ee
\noindent
because when  $\ddo\ge\dve$ then the triangular inequality imposes 
\be
 \dtr\ge(\ddo-\dve)\ .
\ee
In this way the overlap $q$
 (the observable we will study in our Monte Carlo simulations)
is forced  from the triangular inequality to satisfy the relation
\be
  \qea \ge q \ge 4 \qb - 3 \qea\ .
\ee
\noindent
It is reasonable to choose for our simulations for exemple
\be
  \qb = \frac25 \qea\ .
\label{tr}
\ee
\noindent
With these conditions the triangular inequality imposes that
\be
 - \frac75 \qea\leq q\leq \qea\ \mbox{(DT bound)}\  ,
\ee
\noindent
while the ultrametric one imposes 
\be
   \frac25 \qea\leq q\leq\qea\  \mbox{(UM bound)}\ .
\ee
\noindent

Now the difference between these two situations is  evident.
When N is limited (and little) these inequalities will be violated.
The possible ultrametric structure will completely manifest itself only
in the limit of infinite volume. We will try to understand if 
the more we increase the lattice size the better the 
ultrametric bound is satisfied.

We will use for  $q_{EA}$ the thermodynamic limit 
calculated in a work of  Ciria, 
Parisi and Ritort \cite{4d}; at
 $T=1.4$, $q_{EA}=.54$. So we will take for our simulations
\be
  \qb \simeq .21
\label{qfisso}
\ee
\section{The model}

The studied model in this work is a Ising Spin Glass in four dimensions
in absence of external magnetic field. It is defined within a 4D lattice, 
where the  lattice sites  are individualized from the integer values
of the fourdimensional vector $i$.
In every site is defined a Ising spin ($\sigma_i=\pm 1$).
The couplings  $J_{i,j}$ are distributed in binary way
\be
P(J_{i,j})=\frac{1}{2}\delta(J_{i,j}-1)+\frac{1}{2}\delta(J_{i,j}+1)
\ee
\noindent
so that they assume the values $J_{i,j}=\pm 1$ with the same probability
and their action range is limited to only first neighbouring spins.
The hamiltonian of the system is 
\begin{equation}
H[\sigma]\equiv -\frac{1}{2}\sum_{i,j}J_{i,j}\sigma_i\sigma_j\ .
\end{equation}
From these definitions, if we consider the overlap between two thermalized
replicas of the system
\begin{equation}
Q=\frac{1}{N}\sum_{i=1}^{N}\rho_i\sigma_i\hspace{1.3cm}\ ,
\end{equation}
\noindent
it is possible to calculate easily the overlaps distribution function
\begin{equation}
P(q)=\overline{P_{J}(q)}=\overline{\left<\delta (q-\frac{1}{N}\sum_{i=1}^{N}
\rho_i\sigma_i)\right>}
\end{equation}
\noindent
which is the observable upon which we will concentrate our attention.

\section{The dynamic}
The exploration of the phase space at a particular temperature 
(T=1.4 in our case) is directed principally by two algorhythms 

$\bullet$  Monte Carlo Algorhythm

$\bullet$ Bound Algorhythm  

The Monte Carlo Algorhythm used in our simulations generates a
single-spin-flip Glauber dynamic, while the Bound Algorhythm, the very heart of
the simulations, acts re-examining the Monte Carlo flips on the ground
of a bound imposed on the overlaps.
They work in this way:
One takes three replicas of the system, $S_1, S_2, S_3$, with the same
realization of the disorder but different initial spin configurations.
One builds the overlaps

\be
q_{(12)}=\frac{1}{N}\sum_{i=1}^{N}S_1^i S_2^i\hspace{.5cm}
q_{(13)}=\frac{1}{N}\sum_{i=1}^{N}S_1^i S_3^i\hspace{.5cm}
q_{(23)}=\frac{1}{N}\sum_{i=1}^{N}S_2^i S_3^i 
\end{equation}
\noindent
and fixes the  value of $q_{(13)}$ and $q_{(23)}$ equal to $q^{fix}$
(as before establisched)  and defines the tolerance parameter    
 inside of which one consents same fluctuations
\begin{equation}
q_{(13)}=q_{(23)}=[q^{fix}\pm \varepsilon]\ .
\ee
\noindent
The spin changes proposed from the Monte Carlo Algorhythm that satisfy
this request will be accepted. The others will be rejected.

\nin
Now see in detail how to operate the chosen.
Suppose that in a particular instant at a particular site $i$ the
Monte Carlo Algorhythm proposes to the Bound Algorhythm the spin changes

\be
S_1^i\rightarrow\tilde{S_1^i}\hspace{.5cm}
S_2^i\rightarrow\tilde{S_2^i}\hspace{.5cm}
S_3^i\rightarrow\tilde{S_3^i}\ .
\ee
\noindent
Let be \hspace{.18cm}
 $q^V_{(13)}=\frac{1}{N}\sum_{i=1}^{N}S_1^i S_3^i$  \hspace{.18cm}
and \hspace{.18cm} $q^V_{(23)}=\frac{1}{N}
\sum_{i=1}^{N}S_2^iS_3^i$ \hspace{.18cm} the overlaps
calculated from the previous accepted configurations.
\noindent
Consider to begin the spin change $S_1^i\rightarrow\tilde{S_1^i}$.
One calculates the product ${q_{13}^{i}}'\equiv\tilde{S_1^i}S_3^i$
and  tries to replace $q_{13}^i\equiv S_1^i S_3^i$ with ${q_{13}^{i}}'$
in $q^V_{(13)}$

\be
q^{test}_{(13)}\equiv q^V_{(13)}-q_{13}^i+{q_{13}^{i}}'.
\ee
\noindent
Only in the case in which
  $q^{test}_{(13)}\in [q^{fix}-\varepsilon, q^{fix}+\varepsilon]$
we accept the configuration $\tilde{S_1^i}$. 

\nin
Consider now the spin change $S_2^i\rightarrow\tilde{S_2^i}$ and
one imposes an argument similar 
to the previous for the overlap $q^V_{(23)}$.
Also in this case we accept the change only if 
$q^{test}_{(23)}\in [q^{fix}-\varepsilon, q^{fix}+\varepsilon]$.

One examines to finish the spin change  
$S_3^i\rightarrow\tilde{S_3^i}$.
We calculate \hspace{.1cm} ${q_{13}^{i}}''\equiv\tilde{S_3^i}S_1^i$
 \hspace{.1cm}e\hspace{.1cm} ${q_{23}^i}''\equiv\tilde{S_3^i}S_2^i$
and replace \hspace{.1cm} $q_{13}^i$\hspace{.1cm} with 
\hspace{.1cm} ${q_{13}^{i}}''$ \hspace{.1cm} in \hspace{.1cm} $q^V_{(13)}$
and \hspace{.1cm} $q_{23}^i$\hspace{.1cm} with
\hspace{.1cm}  ${q_{23}^i}''$\hspace{.1cm} in
\hspace{.1cm} $q^V_{(23)}$

\be
q^{test}_{(13)}\equiv q^V_{(13)}-q_{13}^i+{q_{13}^i}''
\ee

\be
q^{test}_{(23)}\equiv q^V_{(23)}-q_{23}^i+{q_{23}^i}''\ .
\ee
\noindent
Only if both of them will be included in the defined interval we
will accept the replacement $S_3^i\rightarrow\tilde{S_3^i}$.\\
For our simulations we have established after several tests that
the better value of the tolerance parameter is $\varepsilon=0.04$
(see section 9).

\section{Initial configurations}
The initial spins configurations are so arranged  (it is not the only 
possible way) :

\nin
1) One generates in a random way the spin configuration of the replica 
$S_1$.

\nin
2) One builds the replica $S_3$ so that 60\% of spins 
are (chosen in a random way) equals to the correspondent
spins of the replica $S_1$ and the others are opposites.

\nin
3) The replica $S_2$ will be builded in a way such that 40\%
of spins (chosen casually) are opposites to those correspondent
of the replica $S_3$ and the others are equals.

\nin
These choices allow to begin the simulations in a random spin 
situation and  from values of 
constrained overlaps 
\be
q_{(23)}=q_{(13)}\simeq 0.2 \ .
\ee
\noindent
$q_{(12)}$  varies with the statistic shows in figure
(\ref{a}).

\section{Free and constrained P(q)}

The first part of this work regards the study of the overlaps distribution 
 P(q) at the temperature T=1.4.
This work is useful for two aims. The first is to find some good
annealing schedules for all lattice sizes (taken successively as   
foundation for the study of the ultrametricity) while the second regards
the understanding of the behaviour of this distribution for $q\simeq 0$,
that is to say P(0). The analysis we have done confirms the results already  
obtained in other works \cite{4d}\cite{4d1};
 it seems that there is a clean indipendence
of P(0) from the size of system in agree with the Parisi's theory.
This situation  seems to confirm the existence of a large number  
of pure states in the glass-phase  that are describable with the theory of
SRSB (Spontaneous Replica Symmetry Breaking).
In the figure (\ref{pqmia}) we  report the results of these simulations. 
For lattices of size L=3 and L=4 the averages over the disorder
 are effected  with
1200 different realizations, for  L=5 with 700 samples 
and for the lattice of size L=6 we have averaged with  350 samples.
%%The annealing schedule with which are made the simulations is resumed in  
%%table (\ref{tav1})
%%where the indicated thermalization steps are referred  to the temperature 
%%$T_{i}$ and increase with a cubic law going from $T_{i}$ to $T_{f}$.
%%with the defined step $\Delta  T$.

As already described in detail, the study of the ultrametricity
are made with a constrained dynamic. We have fixed two of three overlaps built
with three replicas of the system and we have observed the behaviour of the  
not constrained one changing the lattice size.
The triangular inequality , considering the bounds imposed to the dynamic,
tell us that

\be
 - \frac75 \qea\leq q\leq \qea\ \mbox{(DT bound)}\\
\ee
\noindent
while the ultrametric one
\be
   \frac25 \qea\leq q\leq\qea\  \mbox{(UM bound)}\\
\ee
\noindent
($q_{EA}$=0.54). For finite  (and little) N these inequalities will be violated. 
A possible ultrametric structure will reveal oneself completely only
in the thermodynamic limit. The annealing schedule
 used for these simulations 
contains  larger numbers of Monte Carlo Steps  compared with the 
ones used for the free P(q) thermalization. Not only the number of MCS
is on average larger that the previous case, but for sizes L=5, L=6, L=7
 and L=8, the step of the simulated annealing is reduced to one half.
Of course the last operation caused a greater precision  
 in the research  of the equilibrium, but so the simulation times are resulted
clearly longer. This choice (extremely prudent) is operated because, cause 
the originality of the dynamic, it is  not still defined any
thermalization method (however the annealing schedule relative to the 
thermalization of the free P(q) would be more than enough to 
assure the equilibrium of our system).

Figure (\ref{tutti}) shows the result of simulations \cite{nostro}
varying of the lattice size.

The behaviour of this curves point out a clear 
dependence of constrained P(q) from the volume of system. Increasing the lattice 
side we observe that the function tends to assume  such a  shape that 
the ultrametric bound is more and more satisfied. Figure (\ref{tutti})
 shows, with vertical lines, both triangular bound and 

\be
 -0.75 \leq q \leq 0.54 \hspace{.5cm}\mbox{(DT bound)}\ ,
\ee
 ultrametric one
\be
 0.21 \leq q\leq 0.54 \hspace{.5cm}\mbox{(UM bound)}\\ .
\ee
\nin

It can be observed that increasing lattice volume there is a systematic shift 
of the peak toward overlap values consented from the ultrametric bound.
Similar remarks can be made about tails relative to negative overlaps
because they too tend toward values indicated from vertical continuous
lines. So it is possible to deduce a clear tendency for large volumes 
to a  fully ultrametric behaviour of the system.
The study of the variation of the largest overlap ($q_{max}^L$)
with the lattice size \cite{nostro}
   (see figure (\ref{tutti}) consents  moreover 
 to establish
a limit to the peak shift  of constrained P(q). 
Best fits made with the function  

\be
 q_{max}^{L}=q_{max}^{\infty}+\frac{\alpha}{L^{\gamma}}
\ee
\nin
indicate that $q_{max}^{\infty}=0.31\pm 0.09$. It is to say
that  in the thermodynamic limit the constrained P(q) will assume her
largest value  just inside the ultrametric bound.

Further confirmation of the ultrametric behaviour of the system can be 
seen studying the variation of integral (see also \cite{nostro})

\vspace{.5cm}

$I_L=\int_{-1}^{0.21}P_L(q_L)(q_L-0.21)^2dq_L+$

\be
\hspace{3cm} +\int_{0.54}^{1}P_L(q_L)
(q_L-0.54)
^2dq_L
\ee
As figure (\ref{fit1}) shows this  tends to zero in the 
thermodynamic limit as expected.

\section{Thermalization}
\label{therm}

It is very important in a computer simulation to be sure that 
obtained results describe the equilibrium system fluctuations.
The main problem in the research of this condition is that generally
we are  interested to the low temperature behaviour of the system where the
phase space is very complex.
Theorically we should make simulations for a infinite time to be sure
that equilibrium is joined but this is obviously impraticable.
We so make everytime errors whit finite-time simulations, but it is
possible to prove  that these ones are comparables with 
statistical fluctuations.
R.N. Bhatt and A.P. Young \cite{sim1} suggested a thermalization criterion   
that is so resumable:

 For every disorder realization one define two overlaps

\begin{equation}
Q(t)=\frac{1}{N}\sum_{i} S_{i}(t_{0})S_{i}(t_{0}+t)
\end{equation}
\noindent

and

\begin{equation}
Q'(t)=\frac{1}{N}\sum_{i} S_{i}^{(1)}(t_{0}+t)S_{i}^{(2)}(t_{0}+t)\ .
\end{equation}
\noindent
Where $t_{0}$ is the esteemed equilibrium time.
In the $t\rightarrow\infty$ limit the two overlaps have 
the same distribution, but also in a finite-time observation
(inside an error margin) they are convergent.

For small times $Q(t)$ assumes values very near to 1 and the P(q)
function will have a peak just
there. In the same situation $Q'(t)$ will have
(starting from a random spin  distribution)
 a Gaussian shape peaked around
 zero and with amplitude $N^{-\frac{1}{2}}$.
When  we increase  $t$, $Q(t)$ and $Q'(t)$ tend to the same distribution.
We can say to be in equilibrium when Monte carlo statistical errors
are bigger than finite-time simulation errors. In other words
when the two overlap distributions are the same function 
unless statistical errors. 
We report in figure (\ref{pq_free_ter}) results of this method
applied to the lattice with size L=3 at the temperature  T=1.4
(everytime we use  $\tau_{0}=t_{0}$).

Considered the originality of dynamic applied
in these simulations (we have three replicas of the system
and so we can't use the Bhatt-Young method) , a criterion 
of thermalization has not yet been defined.
However, we think that a further confirmation ( thermalizing the free P(q)
we have equilibrium in all phase space and in particular in the small
part we are studying with our dynamic)  of the
joined equilibrium for our  system comes from the study of the constrained P(q) 
for different and successive MCS. Specifically, the test we made 
consists in the control, for a given annealing schedule, of the
behaviour of the distribution function with respectively 
the first third, the second third and the last third of MCS used for the
calculation of statistical averages.
The result relative to size L=5 with 100 samples is shown
 in figure (\ref{unterzo}).
It can be  observed that the three curves are pratically identical.
In fact oscillations between the functions are well inside the errors 
from which are afflicted. This test, made on all lattice size, has given,
with our annealing schedule,  good results comforting
ourself about the joined equilibrium of the system.
This is not obviously the faster method ( we are trying to find something
better) but it is sure and for now this is enough.

\section{The tolerance parameter $\varepsilon$}
We have used a particular attention to the behaviour of the constrained P(q)
varying the tolerance parameter. We have concentred our study 
in the lattice of size L=3 and L=5 performing a lot of simulations in order to
understand the develop of the overlaps distribution function 
 under the imposition of more and more restrictive bounds.

The work on the lattice of size L=3 is effectued with a small statistic
but a large range of tolerance parameter values, vice versa
for the lattice of size L=5 we have used a robust statistic but
a little number of $\varepsilon$.
The two results obtained are just compatibles and show, see
figures (\ref{eps1})(\ref{eps2}), a continuous shifting of the tail
relative to the negative overlaps towards values more and more near to that
imposed from the ultrametric bound. This observation  suggest us
 to choice the smallest possible
 tolerance parameter. There is obviously some restrictions.
Too small tolerance parameter 
 values would impose too restrictive bounds in the simulation of the
lattice L=3 while too  big values could be not enough to force  the 
distribution in a decisive way. Moreover 
when the tolerance parameter is bigger than $1+2/5\qea$  we refind
the free overlaps distribution (see figure (\ref{pqmia})) and all
dynamical spin-flips are accepted from the bound algorhythm.
 When $\varepsilon$ decrease the number of spin-flips accepted
 begin smaller and to have enough statistic we must increase 
 the number of MCS for every sample. So we also have to choose
 $\varepsilon$ so that calculator simulations are not too long in time.
After long tests we have established that best compromise in our case
is $\varepsilon=0.04$.  
 
\section{\bf The constrained P(q) $2^{a}$ part}
The evidence of first results  lead us to deepen the feature of the
 ultrametric structure in the examined system. So we further modify
the dynamic without changing the philosophy of the approach.
Now we  fix the overlap
$q_{23}$ and $q_{13}$ to different values (in the first part of the work we
fixed $q_{13}=q_{23}\simeq\frac{2}{5}q_{EA}$).
In particular we made our  simulations with  
\be
 q_{23}\simeq\frac{4}{5}q_{EA}\hspace{1cm} q_{13}\simeq\frac{1}{5}q_{EA}
\ee
\nin
or better $q_{23}=0.43$ and $q_{13}=0.10$.
By  this modification we expect, differently from 
the previous case, that increasing
the size of the lattice the third overlap $q_{12}$
(following the same notation defined in section 5, $q_{12}=Q$) assumes
exclusively the value $q=q_{13}$ how imposed from the ultrametric bound so that
the constrained distribution function of the overlaps tends to a 
Dirac Delta function peaked around  $q_{13}$.

The results of these simulations (see also \cite{nostro})
are shown in figure (\ref{2pp}).
It is evident in this case  a development of the overlaps distribution function
very singular. There is a systematic  shift, starting from small volumes,
 either of the peak of bigger tallness or the smaller peak toward, probably,
the overlap of value $q_{13}=0.10$.
One can see moreover that increasing the lattice size the curve tends
to narrow and the smaller peak is ``absorbed'' in the tail of the
predominant one so that the distribution function assumes the
expected shape.

The study of the biggest overlaps develop with the lattice size
(see figure (\ref{fit2})) consents to establish a limit to the
shift of the predominant peak of constrained P(q).
Our best fits  with the function
\be
 q_{max}^{L}=q_{max}^{\infty}+\frac{\alpha}{L^{\gamma}}
\ee
\nin
affirm that $q_{max}^{\infty}=0.100\pm 0.028$.
So in the infinite size limit our function will tend to settle about a 
value compatible with $q_{13}$.  

To understand at last which is the shape that the constrained P(q) assumes
in the thermodynamic limit we have effected a study of variation
of the integral 
\be
I_L=\int_{-1}^{1} P_L(q_L)(q_L-0.10)^{2}dq_L
\label{qwe}
\ee
with the lattice size. Best fits (see figure (\ref{f3})), always  executed
by the function 
\be
I_L=I^{\infty}+\frac{\alpha}{L^{\gamma}} \  ,
\ee
indicate that
\be
I^{\infty}=\lim_{L\rightarrow\infty}I_L=0.000\pm 0.002 \ ,
\label{infinito}
\ee
so in the thermodynamic limit the integral (\ref{qwe}) will be zero.
In other words, being
\be
P_L(q_L)(q_L-0.10)^{2}\geq 0
\ee
also
\be
I_L\geq0\hspace{1cm} \forall L\  ,
\ee
and in order to satisfy  the (\ref{infinito})  must be verified that
\be
P (q)=\lim_{L\rightarrow\infty}P_L(q_L)=\delta (q_L-0.10)
\ee
as required from the ultrametric hypothesis about the states structure.

\section{\bf Another way to see ultrametric structure}

Another approach we tried to verify the existence of a
 possible ultrametric structure
between the low energy states of a realistic spin glasses  is suggested
from the work \cite{usk2} about the hipercubic lattice .

We take three replicas  $S_1$, $S_2$, $S_3$ of the system with the same
disorder realization and build the  overlaps
$q_{12}$, $q_{13}$, $q_{23}$.

In the previous works we fixed
by constrained dynamic
the value of two of three overlaps (at the same $Q^{fix}$
or at differents  $Q^{fix}$) and  observe the behaviour of the third one.

Now  we define an overlaps bound, say
[$q_{fix1}$, $q_{fix2}$]  (with $q_{fix1}\ge q_{fix2}$),
and then check (with a not constrained dynamic) whether the largest
 overlap falls inside the established bound. In this case
 we calculate
the difference between the other ones $\delta q=q_{mid}-q_{min}$
and  plot the distribution function $P(\delta_q)$. We aspect to
obtain in the termodynamic limit
a Dirac delta function peaked around zero.

We have chosen for our simulations the interval  [$q_{EA},2/5q_{EA}$] for
lattice sizes L=4,6,8 with 400, 200 and 100 samples respectively at the
same temperature T=1.4.

As we can see in figure (\ref{o1}) increasing the volume
 of the lattice the curve tends to narrow more and more showing the presence
of a not trivial structure between the low energy states of the model. As well
as in  the work about hypercubic celles, the behaviour of the curve seems
to manifest a ultrametric feature of the system.

Our last simulation with all others developed in this work we lead to suppose
that the spontaneous
 replica symmetric breaking is a good analitic approach for
the study of realistics Eduards Anderson Spin Glasses.

\section*{\bf Conclusions}
In this work we describe in a very detailed way the constrained
dynamic used in Monte carlo simulation
 of a four dimensional
Eduards Anderson Ising Spin Glass to understand
  the nature of the glass phase
for these systems. We have concentrated our efforts
in the study  of the low energy states structure.
After having defined a metric
 (section 3) in the overlaps space, we have effected
a long series of simulations (section  7, 8, 9, 10) 
in  4D lattice of sizes (L=3,4,5,6,7,8) looking for some confirmations
about a possible analogy with the model SK.
From the obtained results, it seems to be possible to affirm that
a not trivial low energy  states structure in the glass phase
of a  finite dimensional Spin Glass really
exists.
In particular the results shown  in figures (\ref{pqmia}), (\ref{2pp})
 and  (\ref{tutti}) claim the concreteness of the hypothesis proposed by 
G. Parisi about the origin and the kind of this structure.
The origin should be researched in the spontaneous replica symmetry breaking 
\cite{Parisi1}\cite{Parisi3}\cite{Parisi4}\cite{Ord_par}
 and the kind of the  structure really 
seems to be the ultrametric one as already
established for the SK model.

\pagebreak

\pagebreak

\f
\centerline{
\epsfig{figure=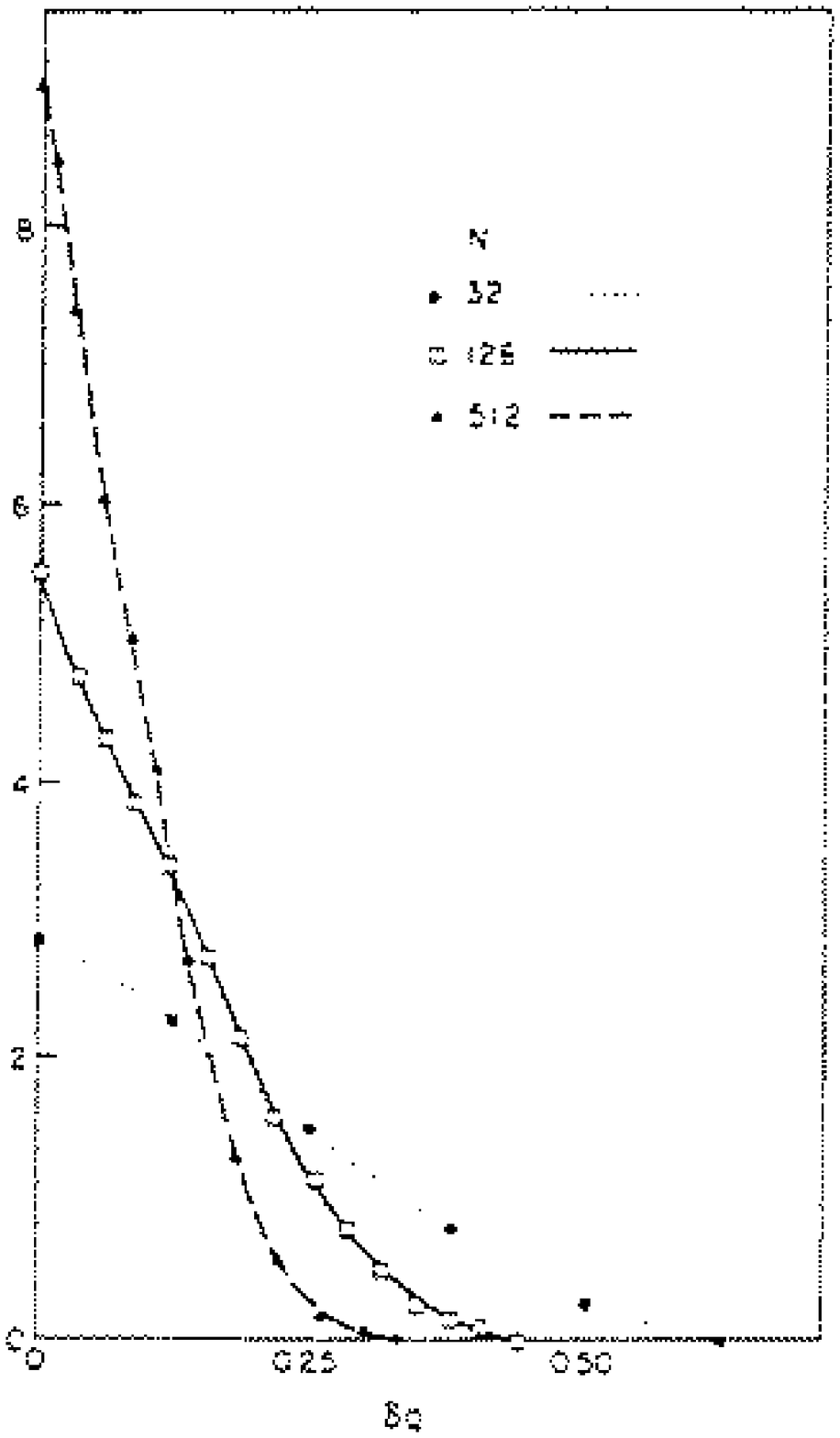,bbllx=50pt,bblly=98pt,bburx=545pt,
bbury=600pt,width=300pt,height=240pt}}
\caption[109]{Probability distribution of the difference between
the two smaller overlaps ($\delta q=q_{mid}-q_{min}$) when the bigger
overlap is smaller than 0.4 from $q_{EA}$
 in a four dimensional Ising Spin Glass
for different lattice sizes \cite{4d}}
\label{u4d}
\ef

\vspace{1cm}
\f
\centerline{
\epsfig{figure=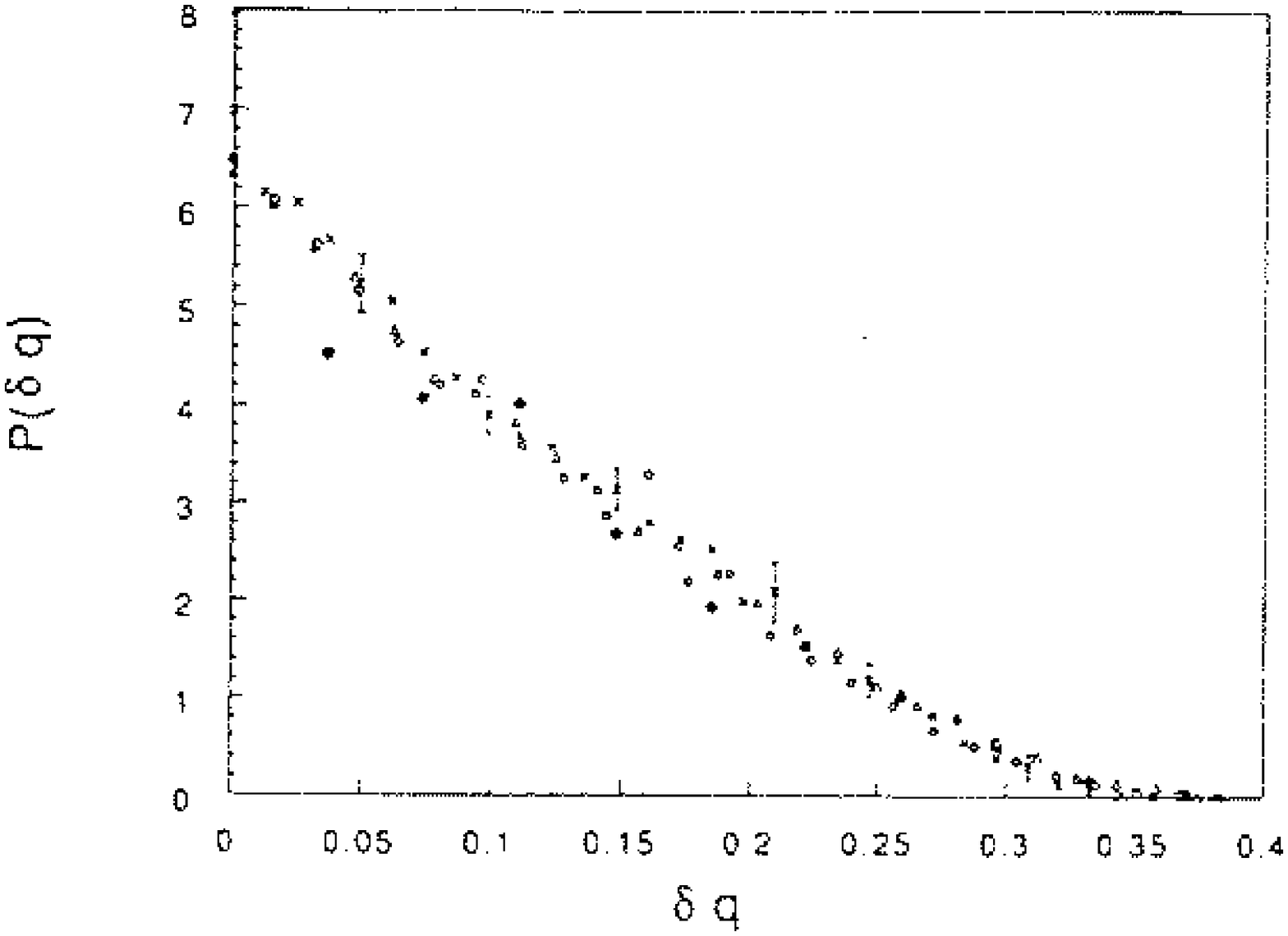,bbllx=50pt,bblly=98pt,bburx=545pt,
bbury=600pt,width=300pt,height=240pt}}

\vspace{.8cm}
\caption[158]{Probability distribution of the difference between the two 
smaller overlaps ($\delta q=q_{mid}-q_{min}$) for a fixed larger overlap
 value q=0.5. The temperature is $0.6T_{crit.}$. In the $N\rightarrow\infty$
limit, the distribution is expected to become a $\delta$ function
at the  origin  \cite{sim1}}
\label{pqmid12}
\ef
\begin{figure}
\centerline{
\epsfig{figure=
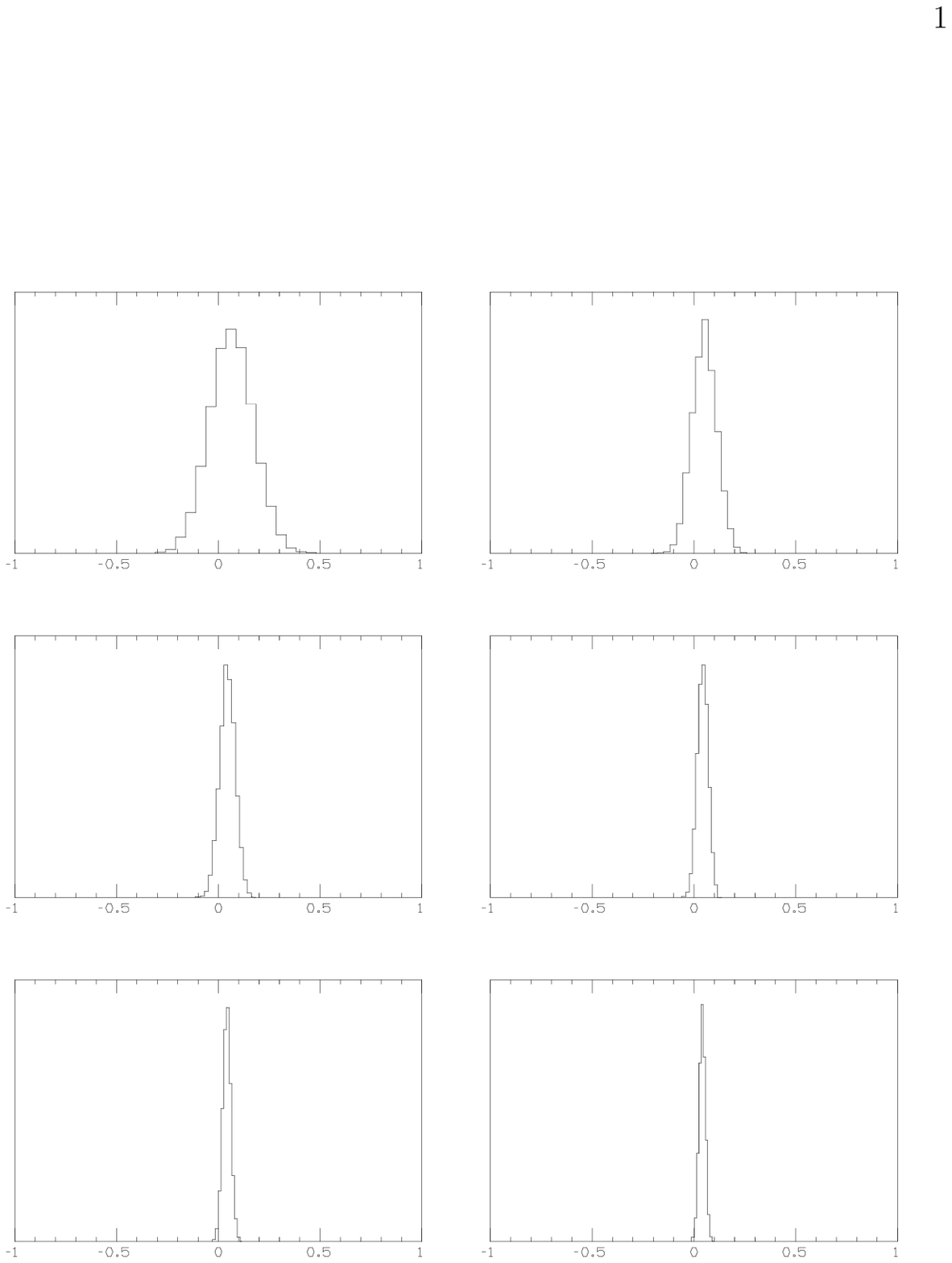,bbllx=50pt,bblly=200pt,bburx=449pt,
bbury=649pt,width=500pt,height=500pt}
        }

\vspace{-.5cm}

\caption[125]{Distribution probability P($q_{12}$) (in y-axes)
 of starting values that $q_{(12)}$ (in x-axes)
assumes varying the initial  spin configurations with 10000 different
starts for sides L=3, 4, 5, 6, 7, 8 in reading order}
\label{a}
\end{figure}

\begin{figure}
\centerline{
\epsfig{figure=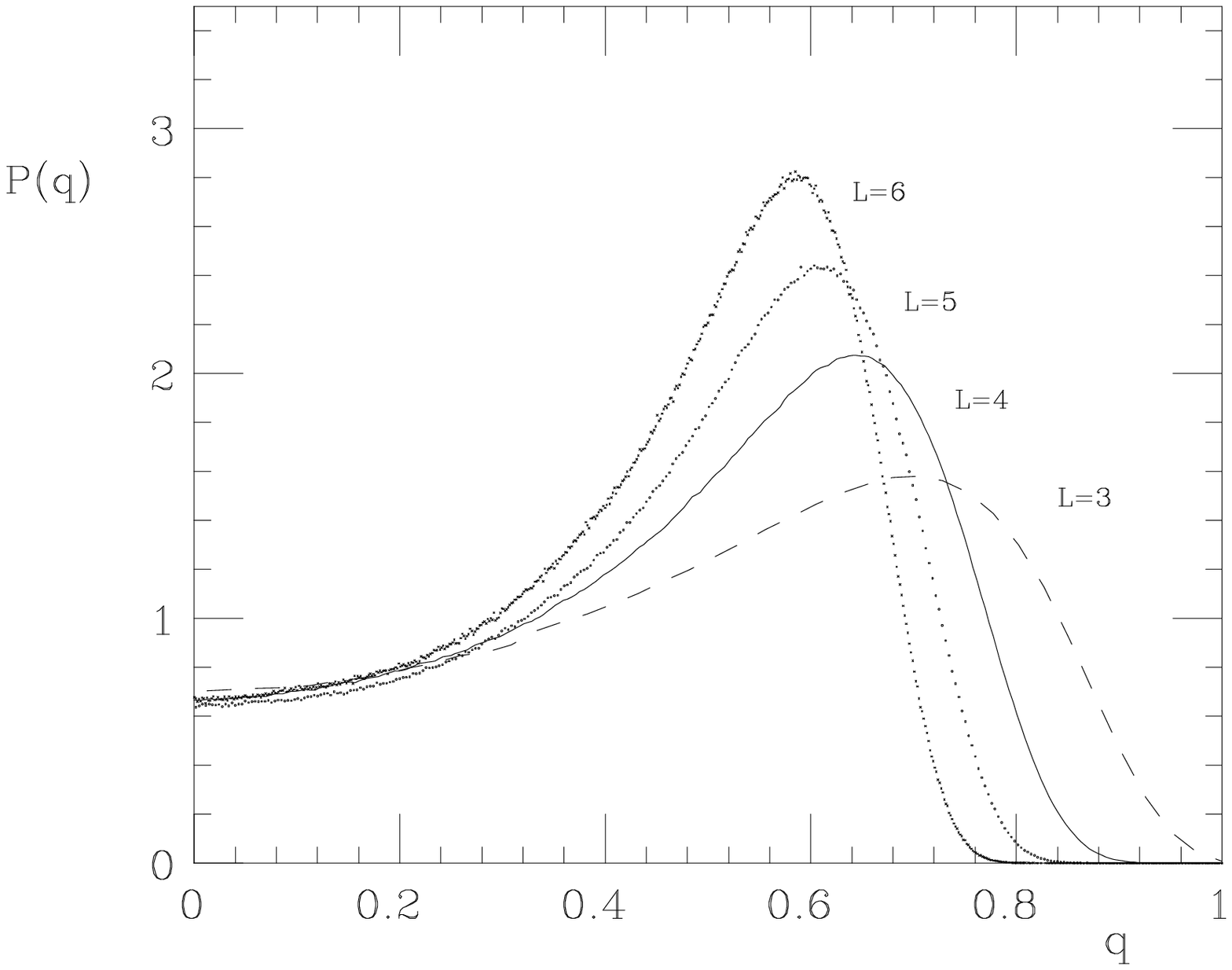,bbllx
=50pt,bblly=98pt,bburx=545pt,
bbury=600pt,width=300pt,height=280pt,angle=90}}

\vspace{-2cm}

\caption{Distribution probability of the overlaps 
 P(q) at T=1.4 for different sizes of the lattice L=3, 4, 5, 6}
\label{pqmia}
\centerline{
\epsfig{figure=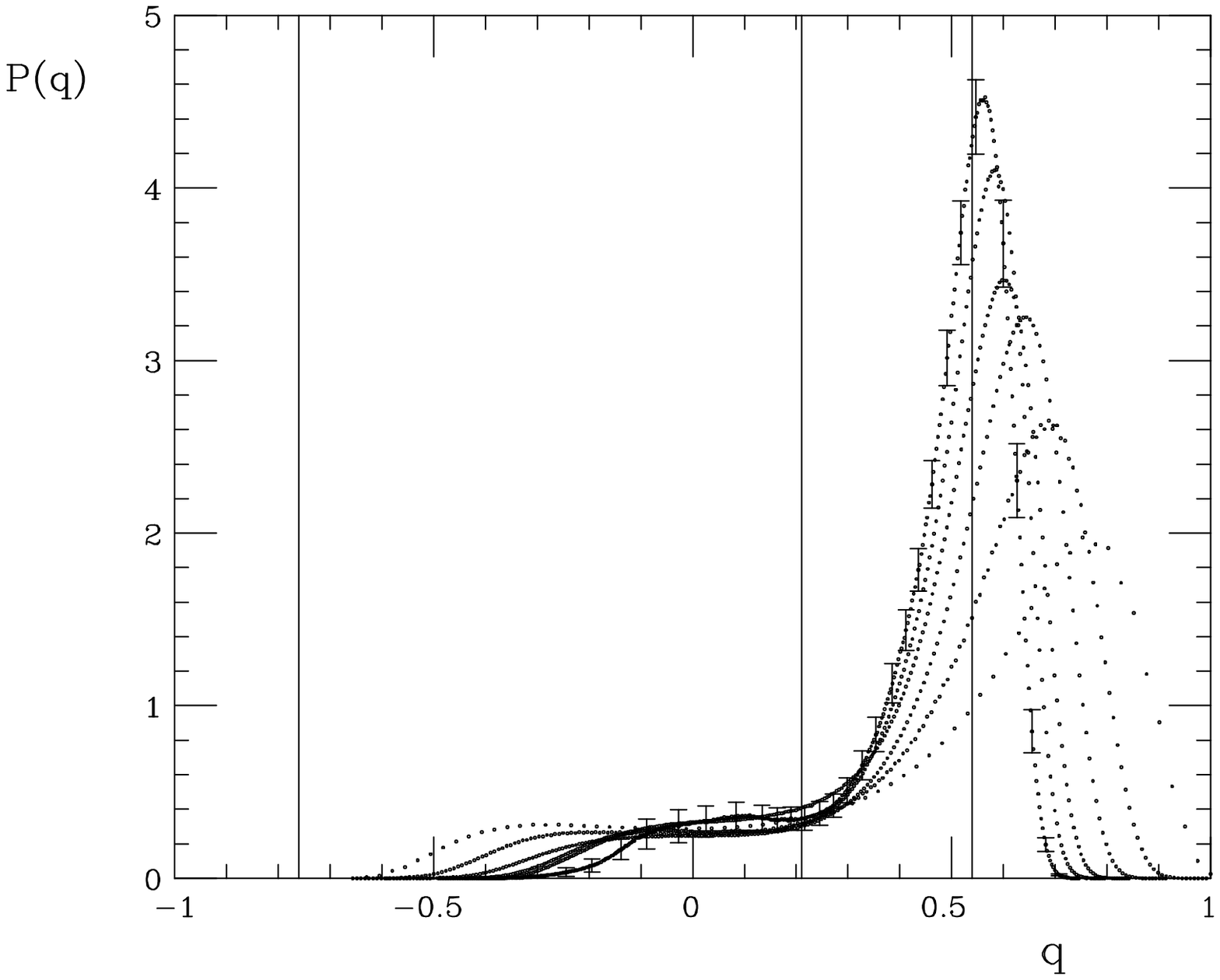,bbllx=50pt,
bblly=98pt,bburx=545pt,bbury=600pt,width=300pt,height=280pt,angle=90}
        }

\vspace{-2cm}

\caption[87]{Evolution of the probability distribution of overlaps P(q)
 with a constrained dynamic a T=1.4 for lattice size
 L=3, 4, 5, 6, 7, 8.}
\label{tutti}
\end{figure}

\begin{figure}
\centerline{
\epsfig{figure=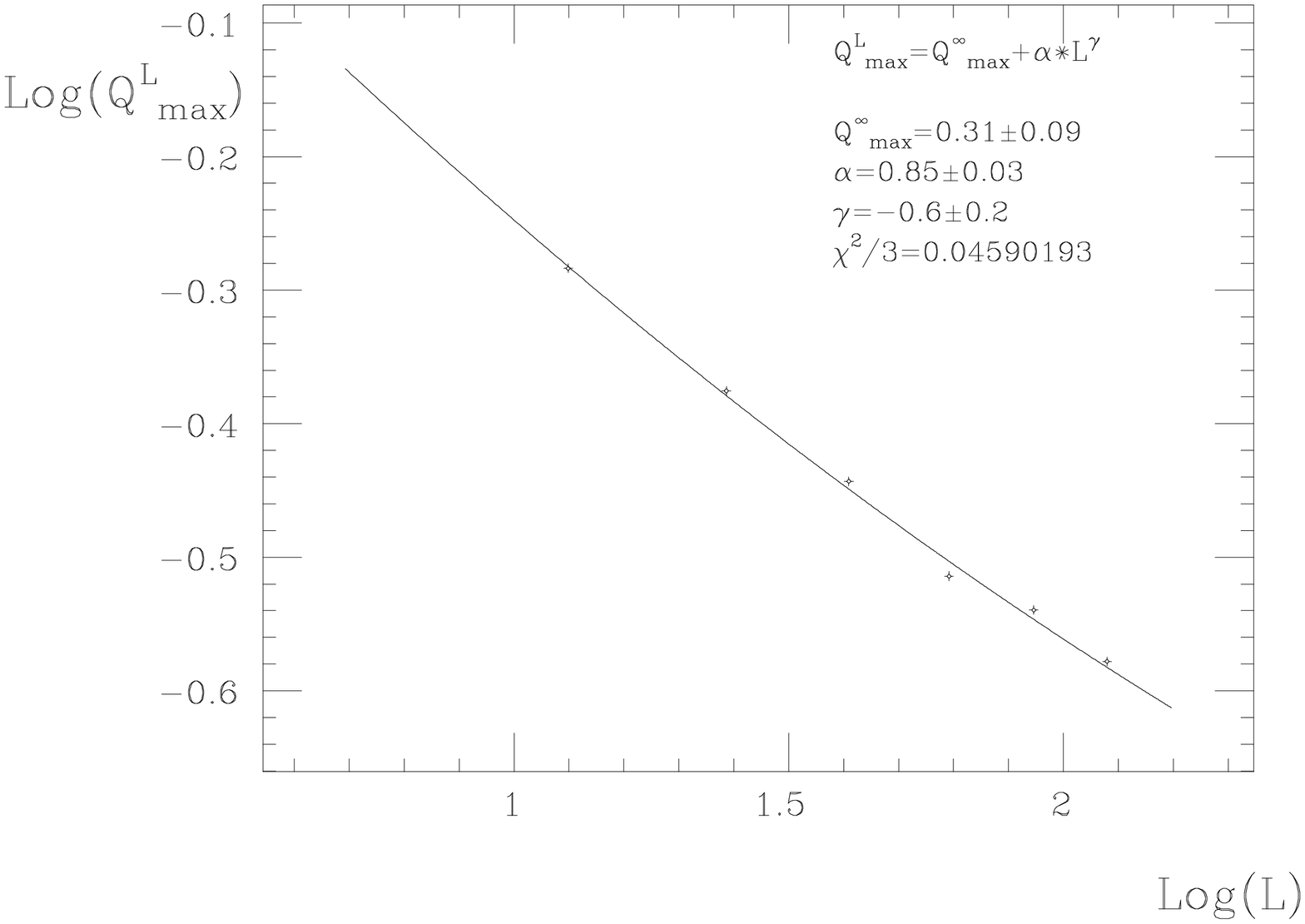
,bbllx=50pt,bblly=98pt,bburx=545pt,
bbury=600pt,width=300pt,height=240pt,angle=90}}

\vspace{-.7cm}
\caption[986]{Fit of biggest overlaps for the constrained P(q) with the
 lattice size by the function
$I^{L}_{max}=I^{\infty}_{max}+\frac{\alpha}{L^{\gamma}}$ in log-log scale}
\label{fit1}
\centerline{
\epsfig{figure=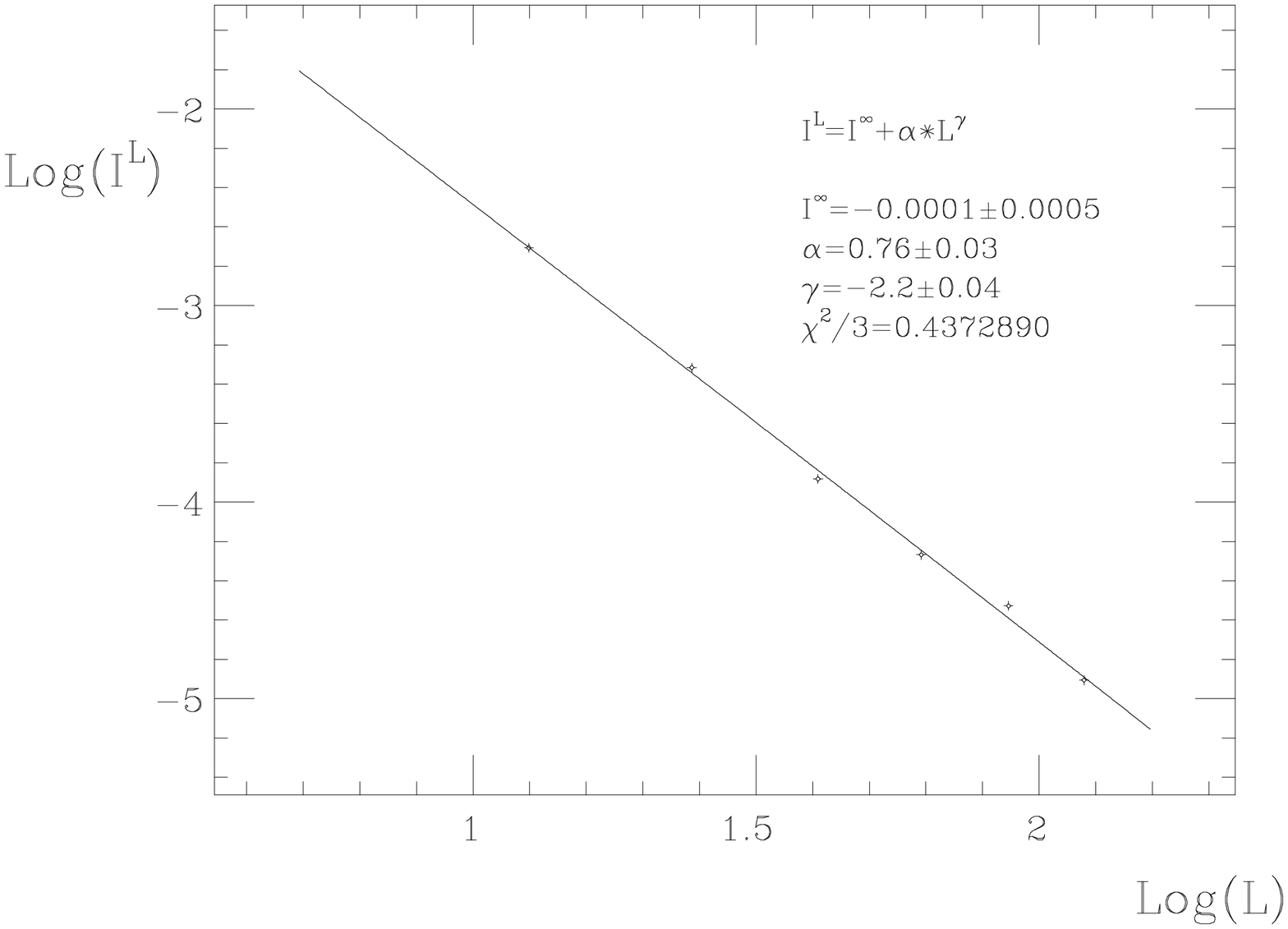
,bbllx=50pt,bblly=98pt,bburx=545pt,
bbury=600pt,width=300pt,height=240pt,angle=90}}
\label{area}

\vspace{-.7cm}
\caption[13]{Evolution of the integral
$I_L=\int_{-1}^{0.21}P_L(q_L)(q_L-0.21)^2dq_L
+\int_{0.54}^{1}P_L(q_L)(q_L-0.54)
^2dq(L)$
with the  lattice size
in log-log scale}
\end{figure}

\begin{figure}[hbt]
\centerline{
\epsfig{figure=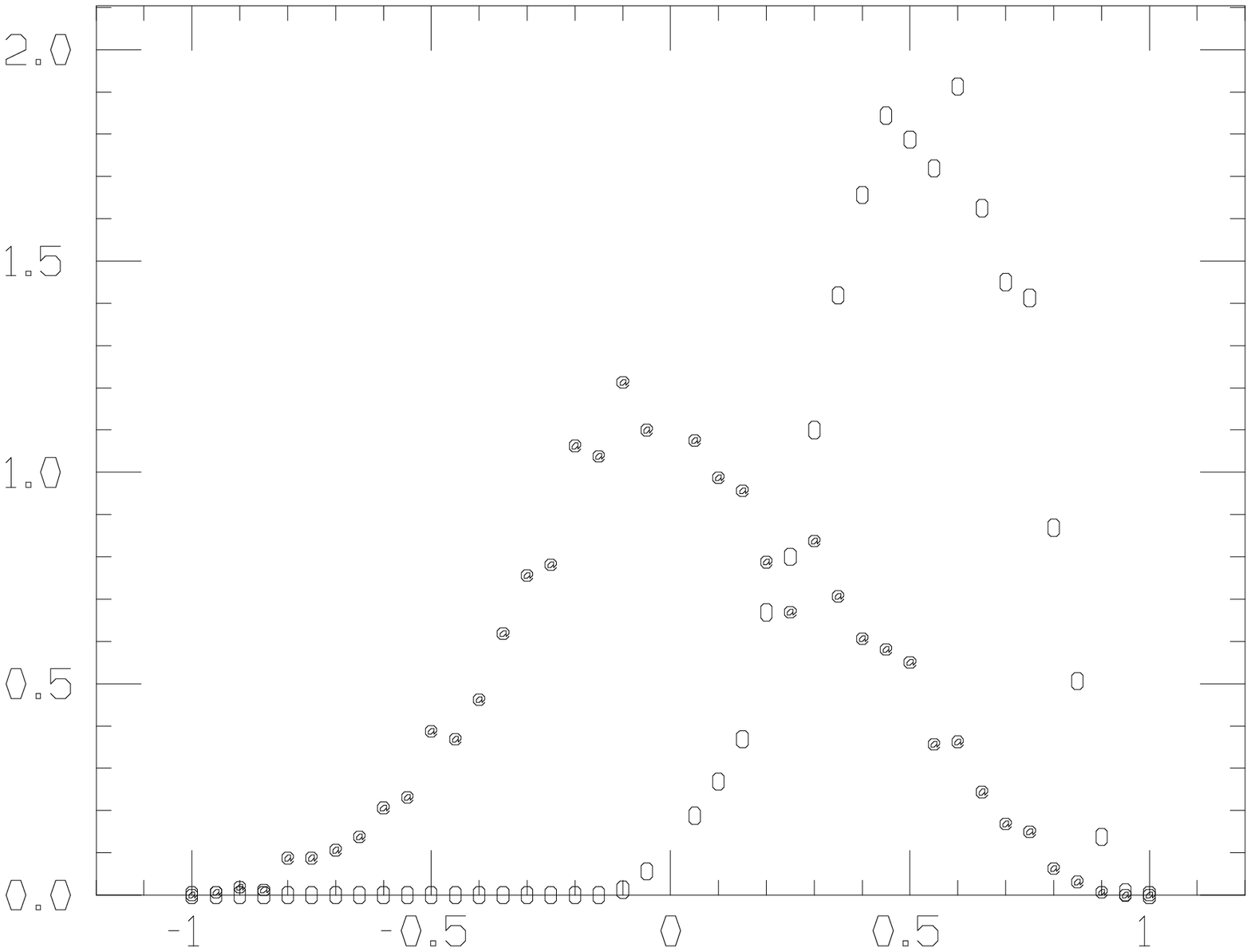,bbllx=50pt,bblly=98pt,bburx=545pt,%
bbury=600pt,width=4.5cm,height=4.5cm,angle=90}
\hspace{.3cm}
\epsfig{figure=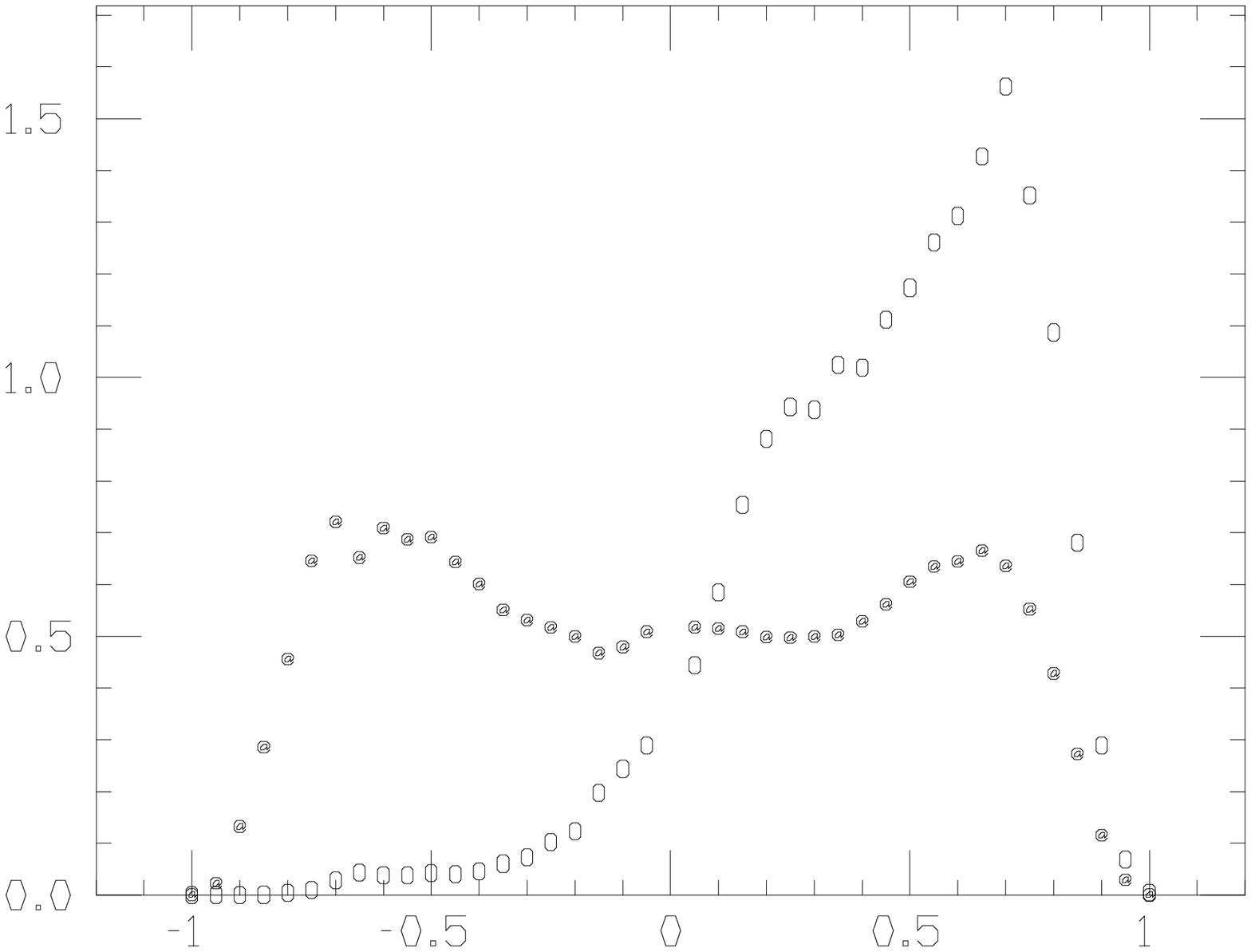,bbllx=50pt,bblly=98pt,bburx=545pt,%
bbury=600pt,width=4.5cm,height=4.5cm,angle=90}}
\vspace{2cm}
\centerline{
\epsfig{figure=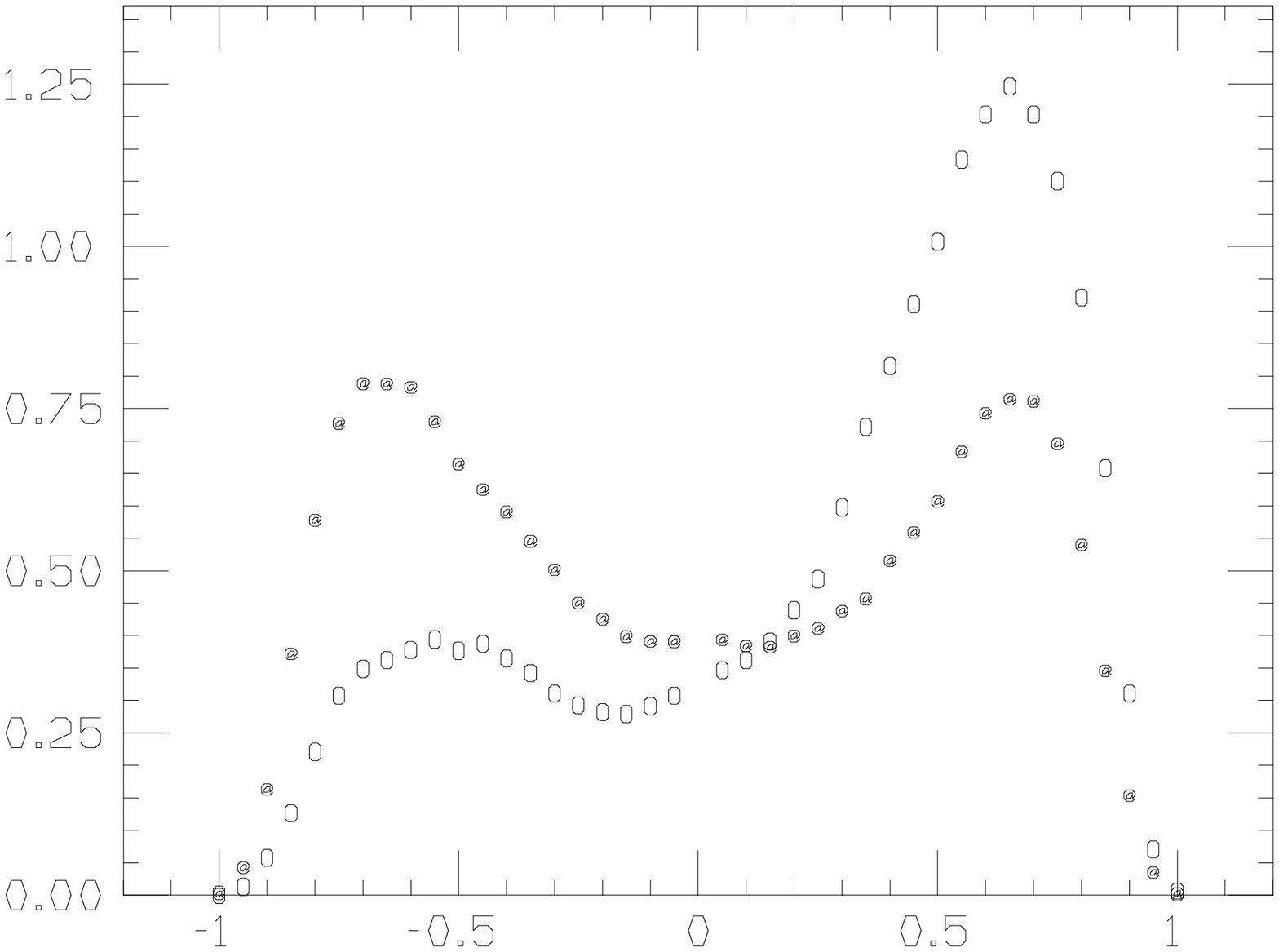,bbllx=50pt,bblly=98pt,bburx=545pt,%
bbury=600pt,width=4.5cm,height=4.5cm,angle=90}
\hspace{.3cm}
\epsfig{figure=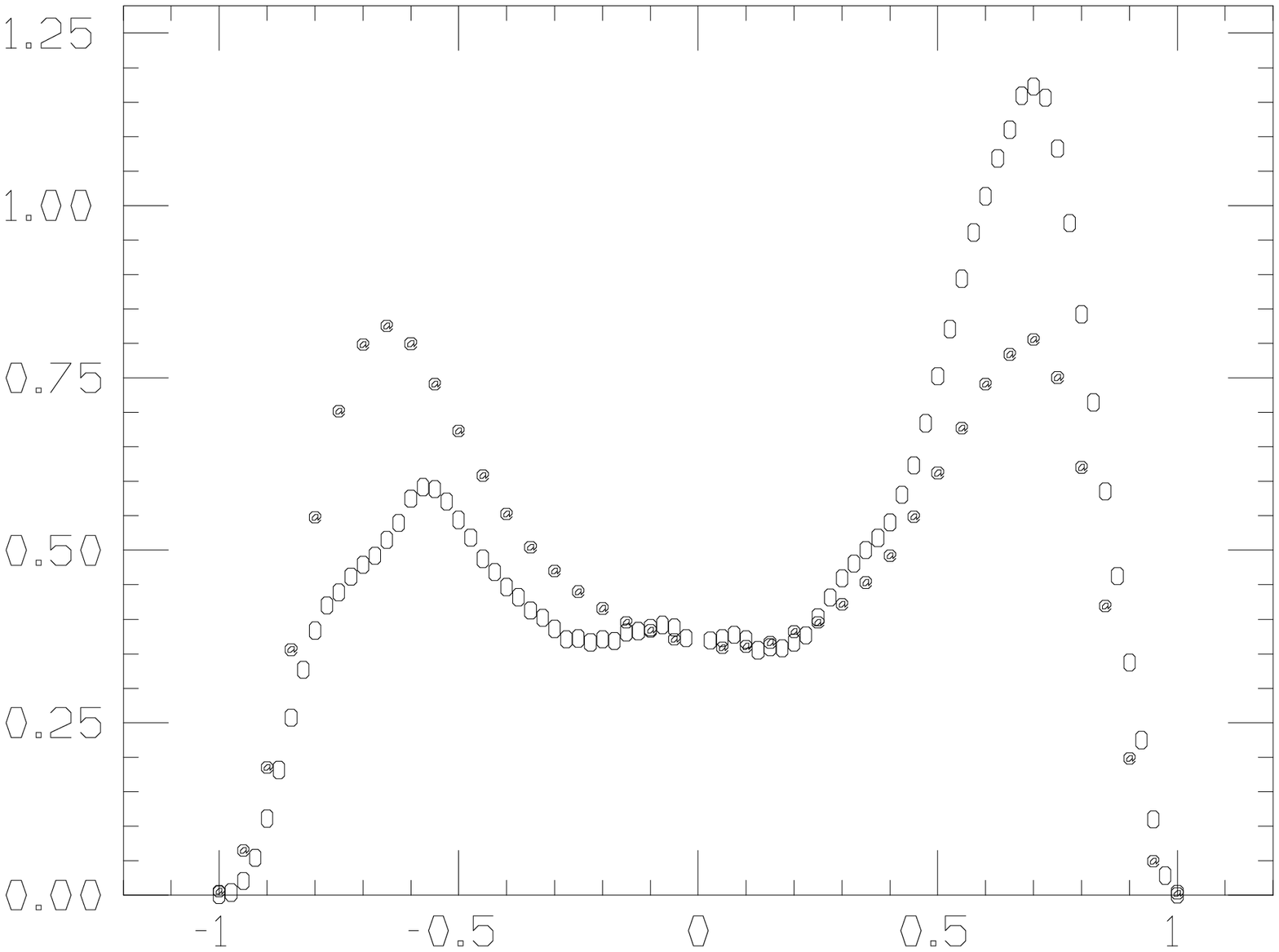,bbllx=50pt,bblly=98pt,bburx=545pt,%
bbury=600pt,width=4.5cm,height=4.5cm,angle=90}
        }
\vspace{2cm}
\centerline{
\epsfig{figure=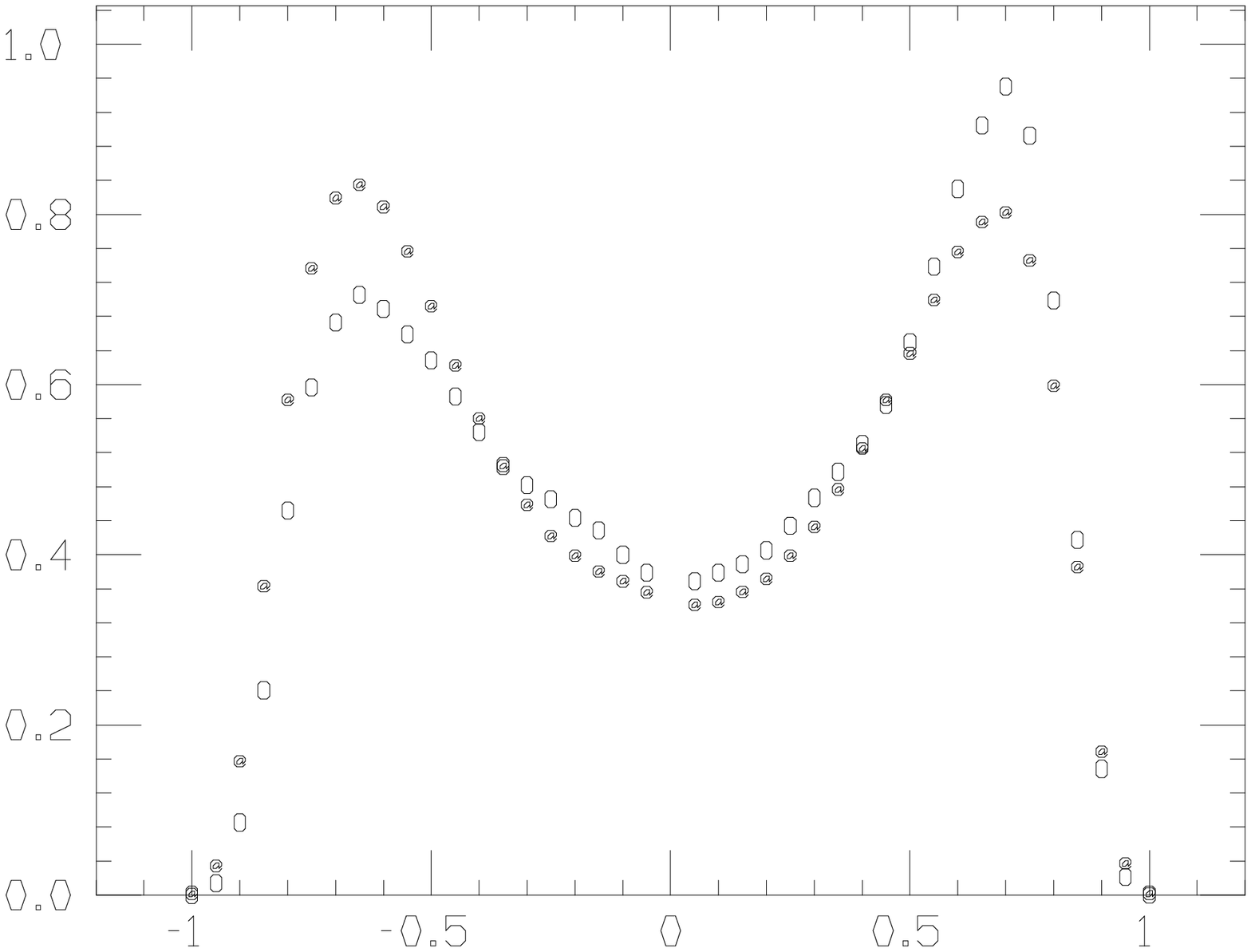,bbllx=50pt,bblly=98pt,bburx=545pt,%
bbury=600pt,width=4.5cm,height=4.5cm,angle=90}
\hspace{.3cm}
\epsfig{figure=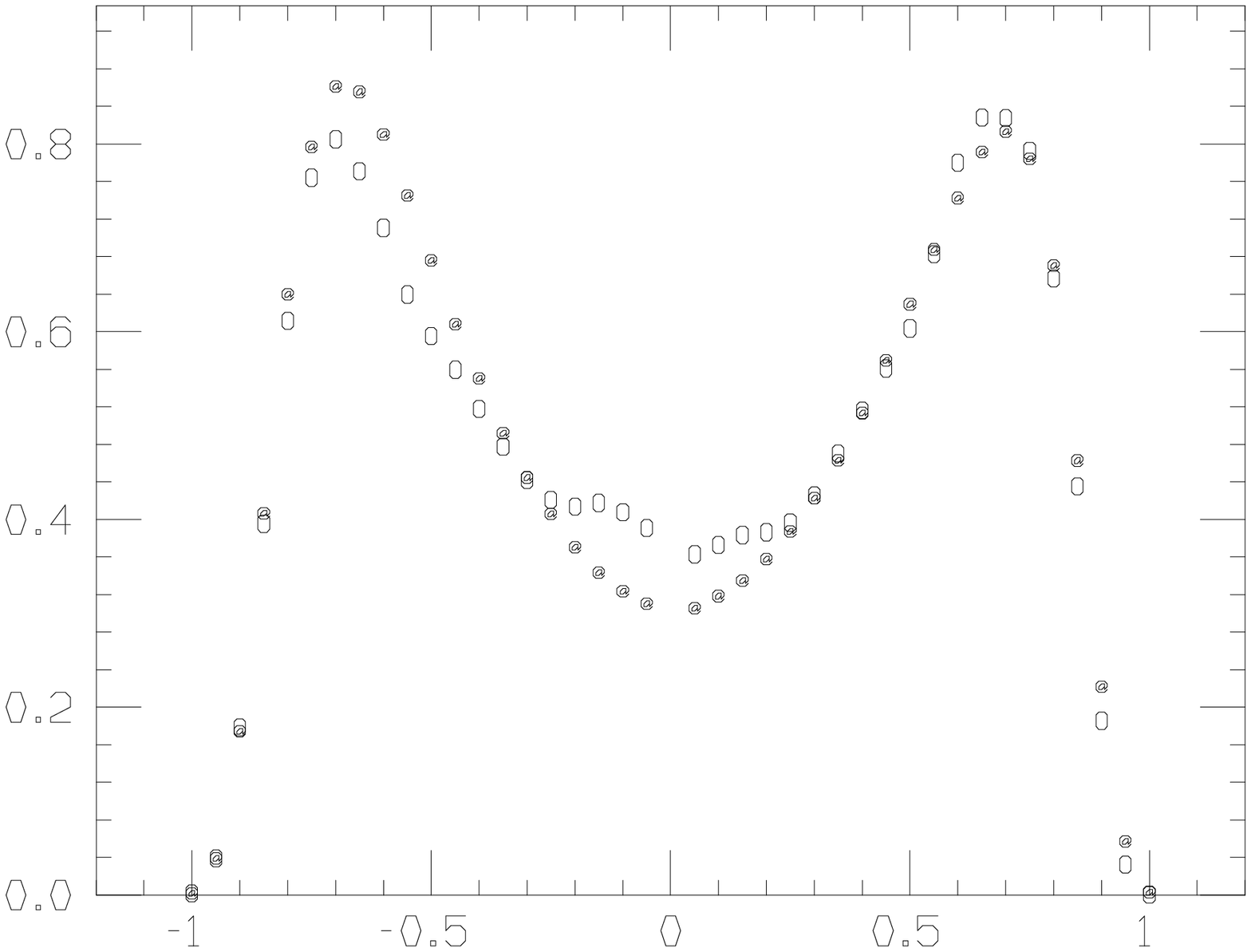,bbllx=50pt,bblly=98pt,bburx=545pt,%
bbury=600pt,width=4.5cm,height=4.5cm,angle=90}
\hspace{.3cm}
\epsfig{figure=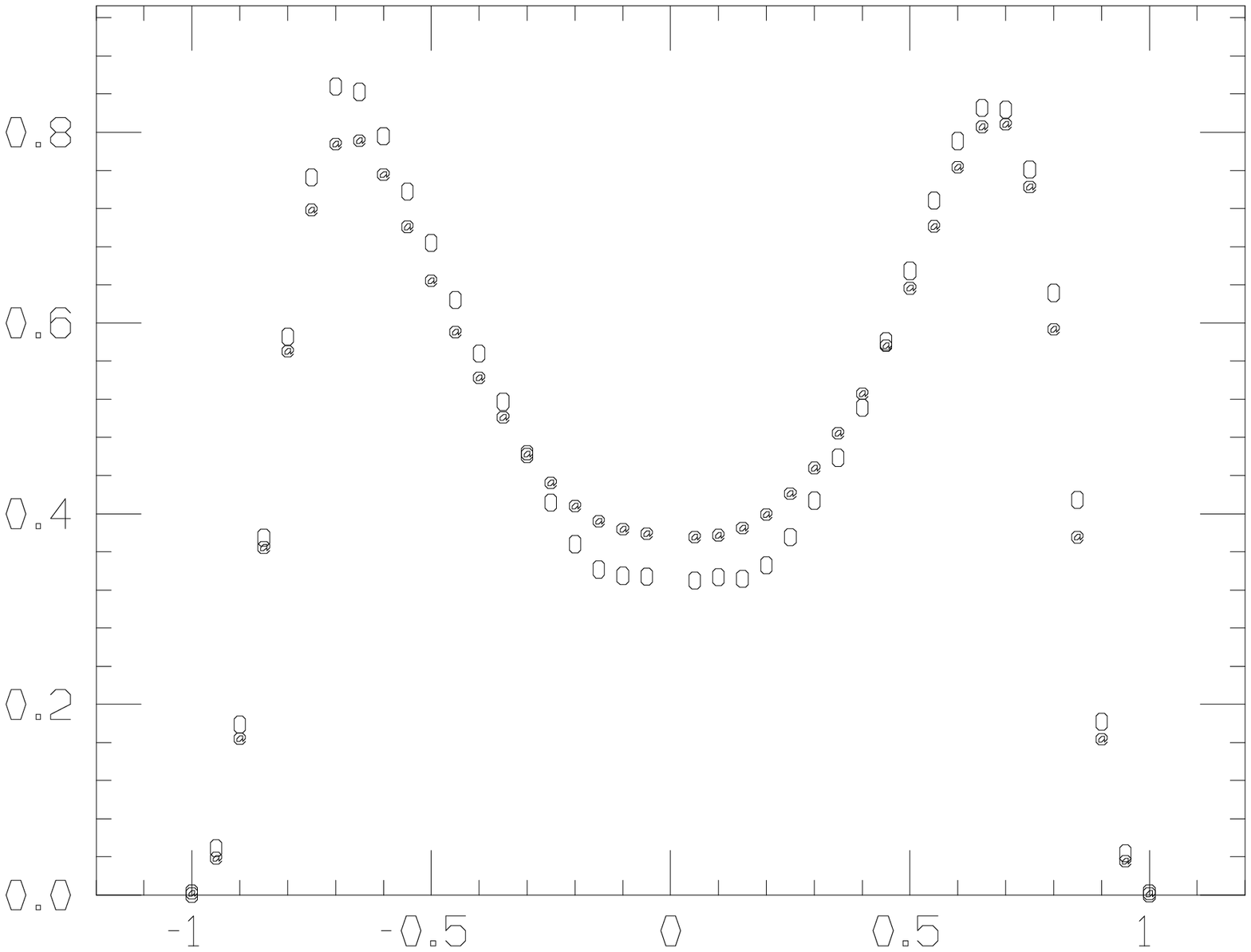,bbllx=50pt,bblly=98pt,bburx=545pt,%
bbury=600pt,width=4.5cm,height=4.5cm,angle=90}
        }

\vspace{-.5cm}
        
\caption{Evolution of free overlap distributions P(q) and P(q') (in y-axes) 
 (see section 
\ref{therm}) as a function of thermalization steps (in x-axes).
$t_{0}=10,10^{2}, 10^{3}, 10^{4}, 2\cdot 10^{4}, 3\cdot 10^{4}, 5\cdot 10^{4}$ 
in  reading order}
\label{pq_free_ter} 
\end{figure}

\begin{figure}
\centerline{
\epsfig{figure=
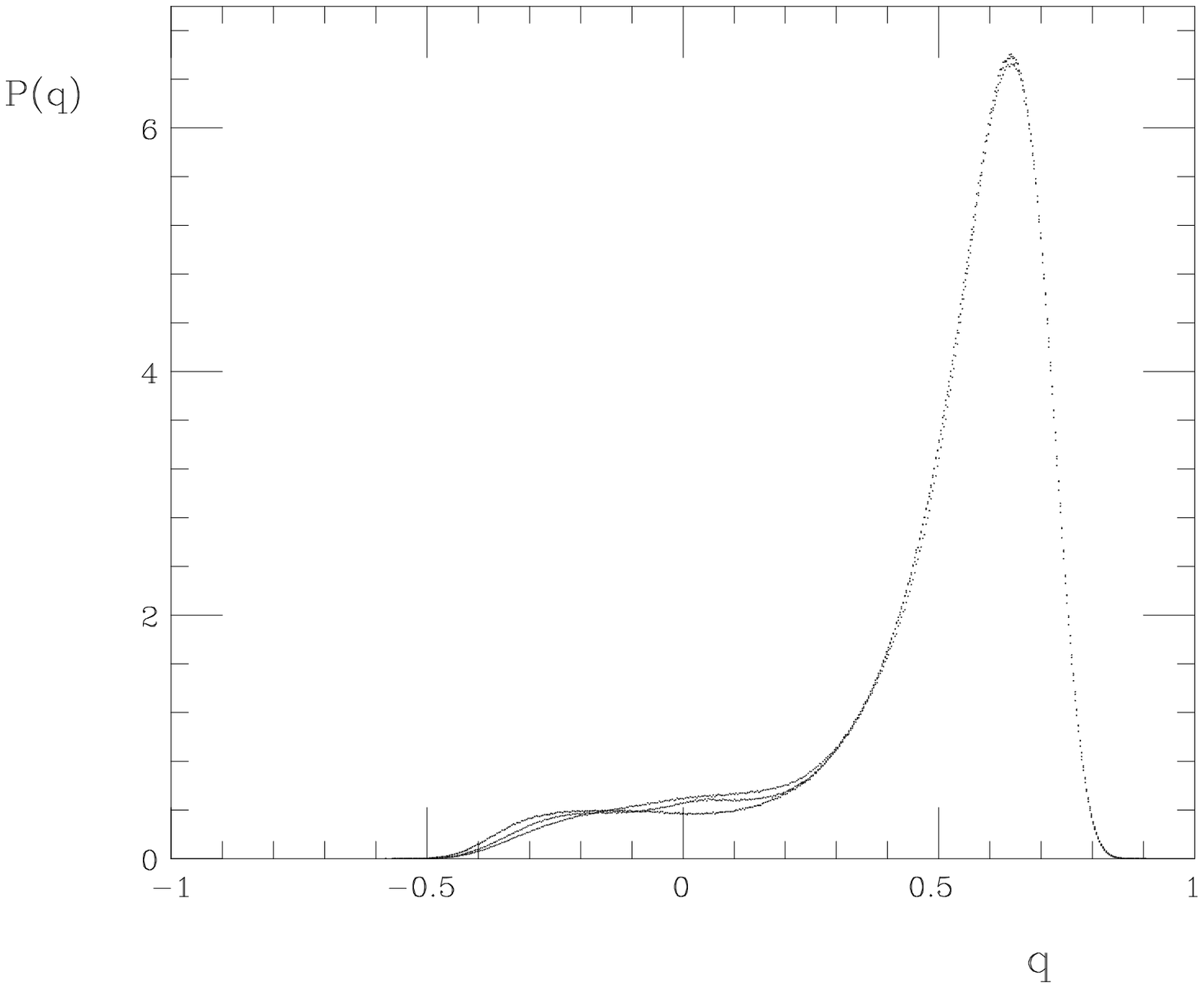,bbllx=50pt,bblly=98pt,bburx=545pt,
bbury=600pt,width=450pt,height=450pt,angle=90}}
\caption[126]{The tree curves (basically coinciding in the plot) are
for constrained P(q) in the first, the second and the third of the run
 after the annealing scheme described in the text), L=5, 100 samples}
\label{unterzo}
\end{figure}

\begin{figure}
\centerline{
\epsfig{figure=
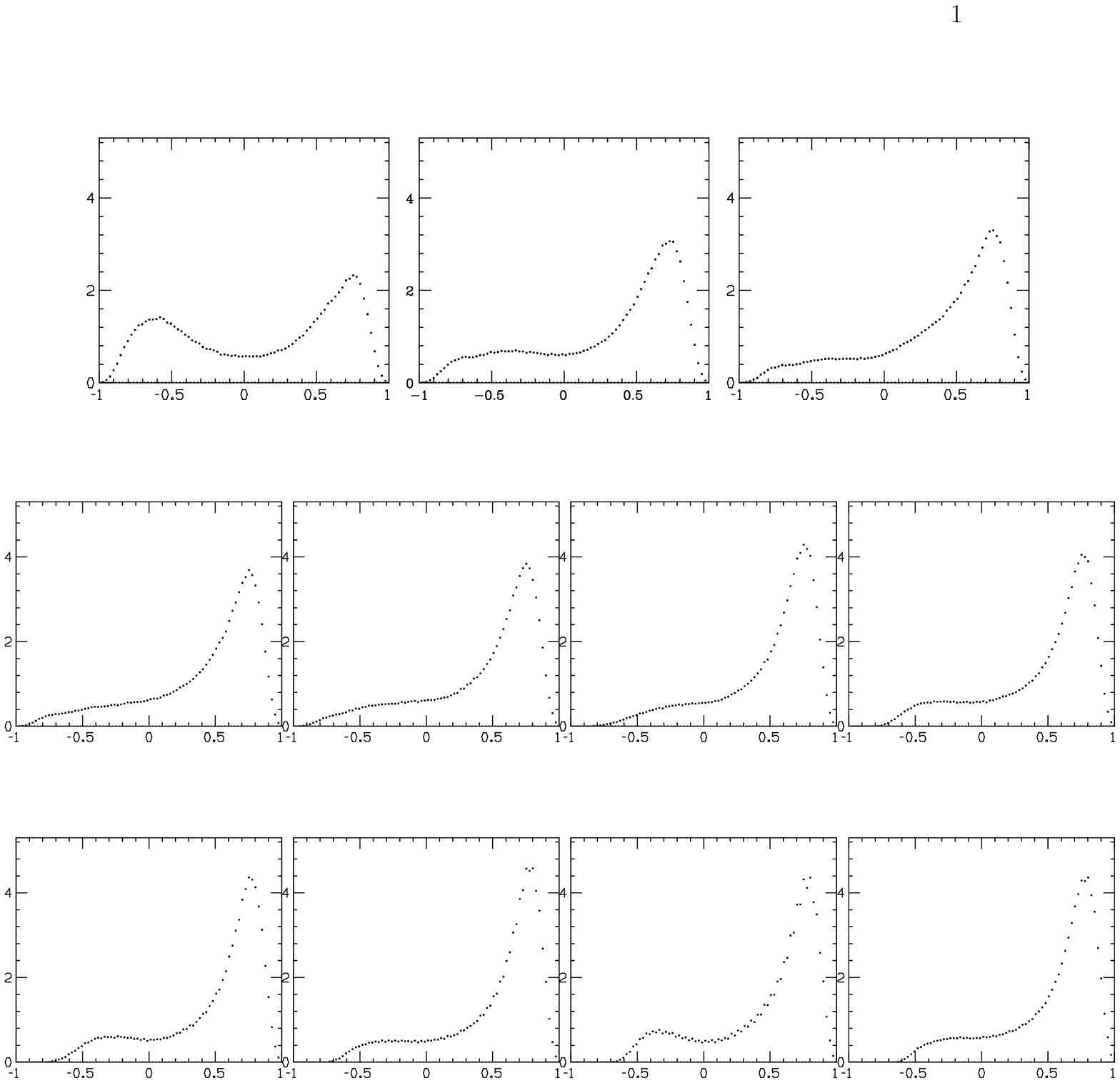,bbllx=50pt,bblly=200pt,bburx=449pt,
bbury=649pt,width=400pt,height=500pt}
            }

\vspace{-.5cm}

\caption{Evolution of the constrained P(q) (in y-axes) as a function of 
overlap (in x-axes) 
  for different values of tolerance
  parameter $\varepsilon= $0.8, 0.6, 0.4,
0.3, 0.2,
 0.15, 0.12, 0.1, 0.08, 0.06, 0.04. L=3, T=1.4, samples=64.}
\label{eps1}
\end{figure}

\begin{figure}
\centerline{
\epsfig{figure=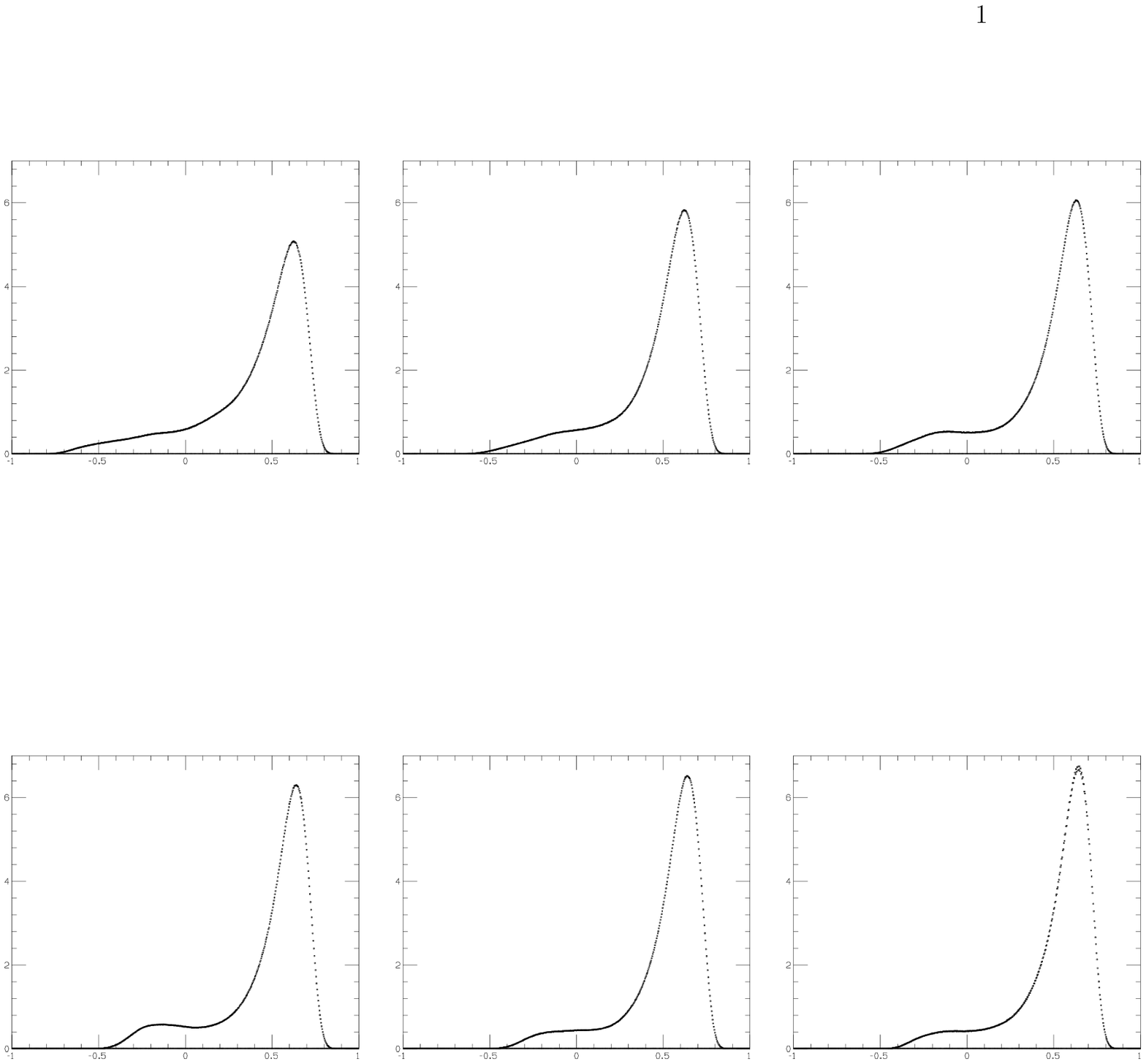,bbllx=50pt,bblly=200pt,bburx=449pt,
bbury=649pt,width=410pt,height=500pt}}

\vspace{-.5cm}

\caption{Evolution of constrained P(q)  (in y-axes) as a function of
overlap (in x-axes) for different values of tolerance  parameter 
  $\varepsilon= $ 0.22,
0.12, 0.09, 0.05, 0.04, 0.02,
 L=5, T=1.4, samples=100.}
\label{eps2}
\end{figure}

\begin{figure}[h]
\centerline{
\epsfig{figure=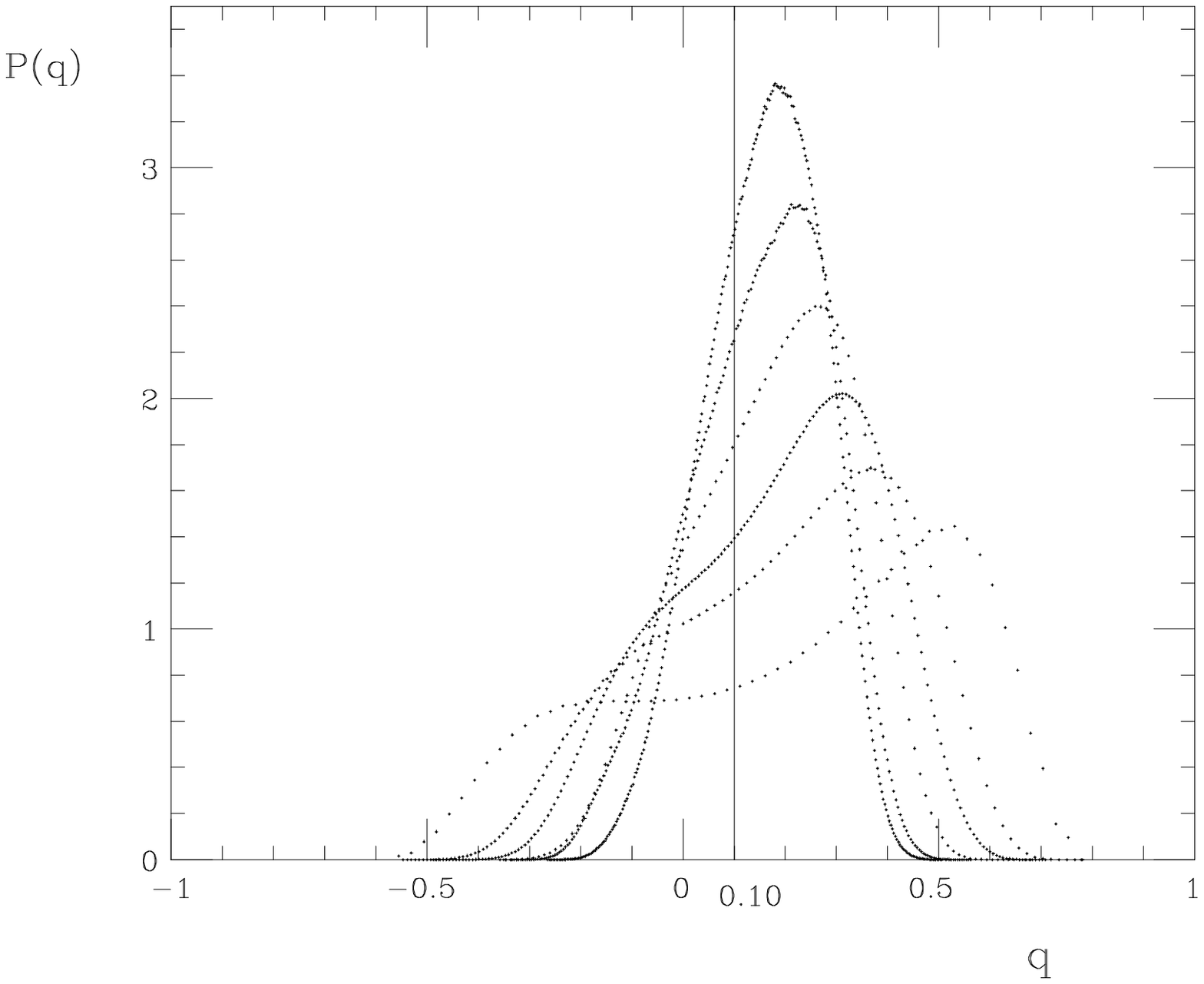,
bbllx=50pt,bblly=98pt,bburx=545pt,
bbury=600pt,width=300pt,height=300pt,angle=90}}
 \caption[87]{Study of the behaviour of constrained P(q) with lattice size;
 L=3, 4, 5, 6, 7, 8. T=1.4 ($q_{min}\neq q_{max}$)}
\label{2pp}
\end{figure}
\begin{figure}[h]
\centerline{
\epsfig{figure=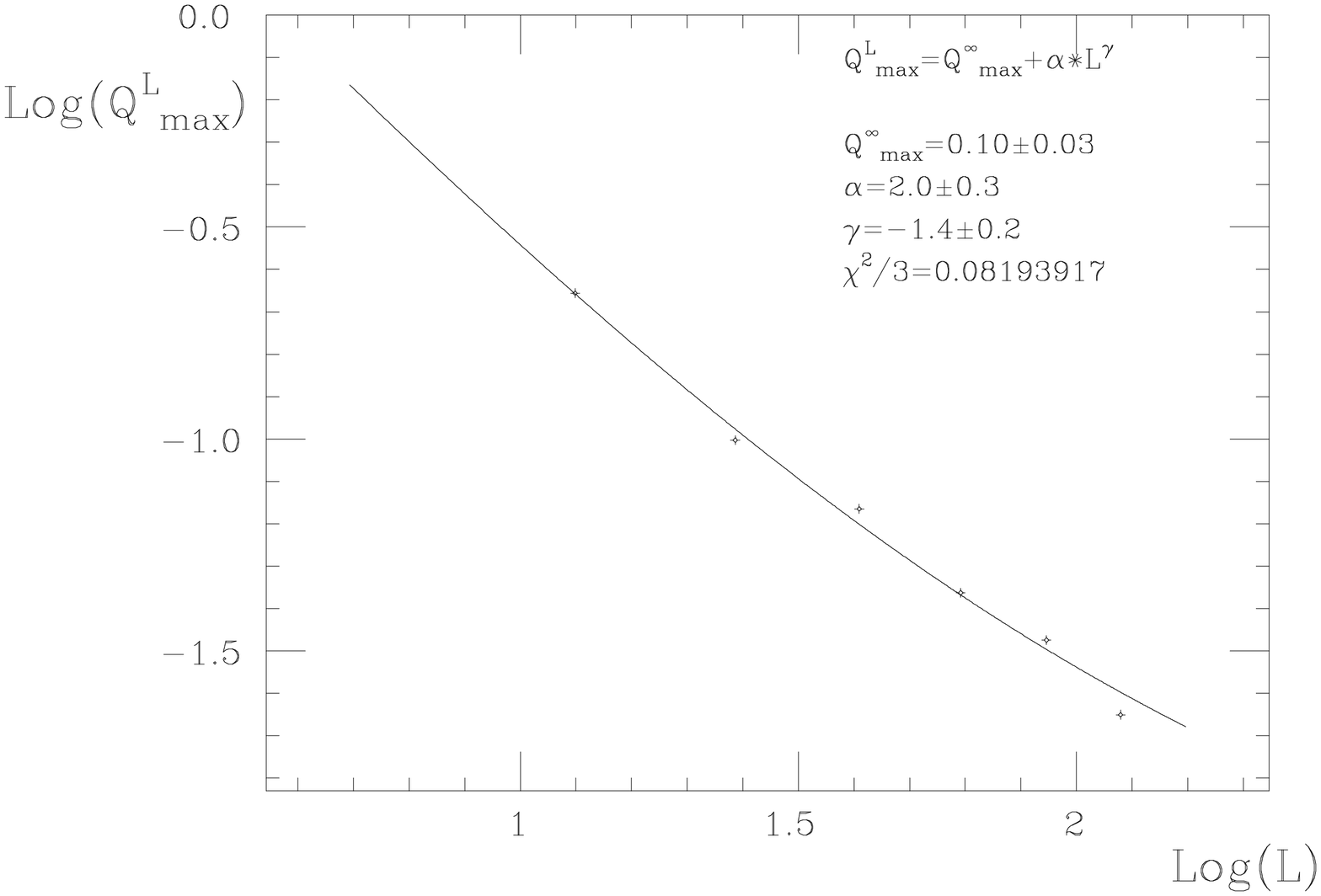
,bbllx=50pt,bblly=98pt,bburx=545pt,
bbury=600pt,width=280pt,height=240pt,angle=90}}

\vspace{-.5cm}
\caption[986]{Fit of the predominant parameter peak of the constrained P(q)
 by the
function
$q_{max}^{L}=q_{max}^{\infty}+\frac{\alpha}{L^{\gamma}}$ in log-log scale}
\label{fit2}
\centerline{
\epsfig{figure=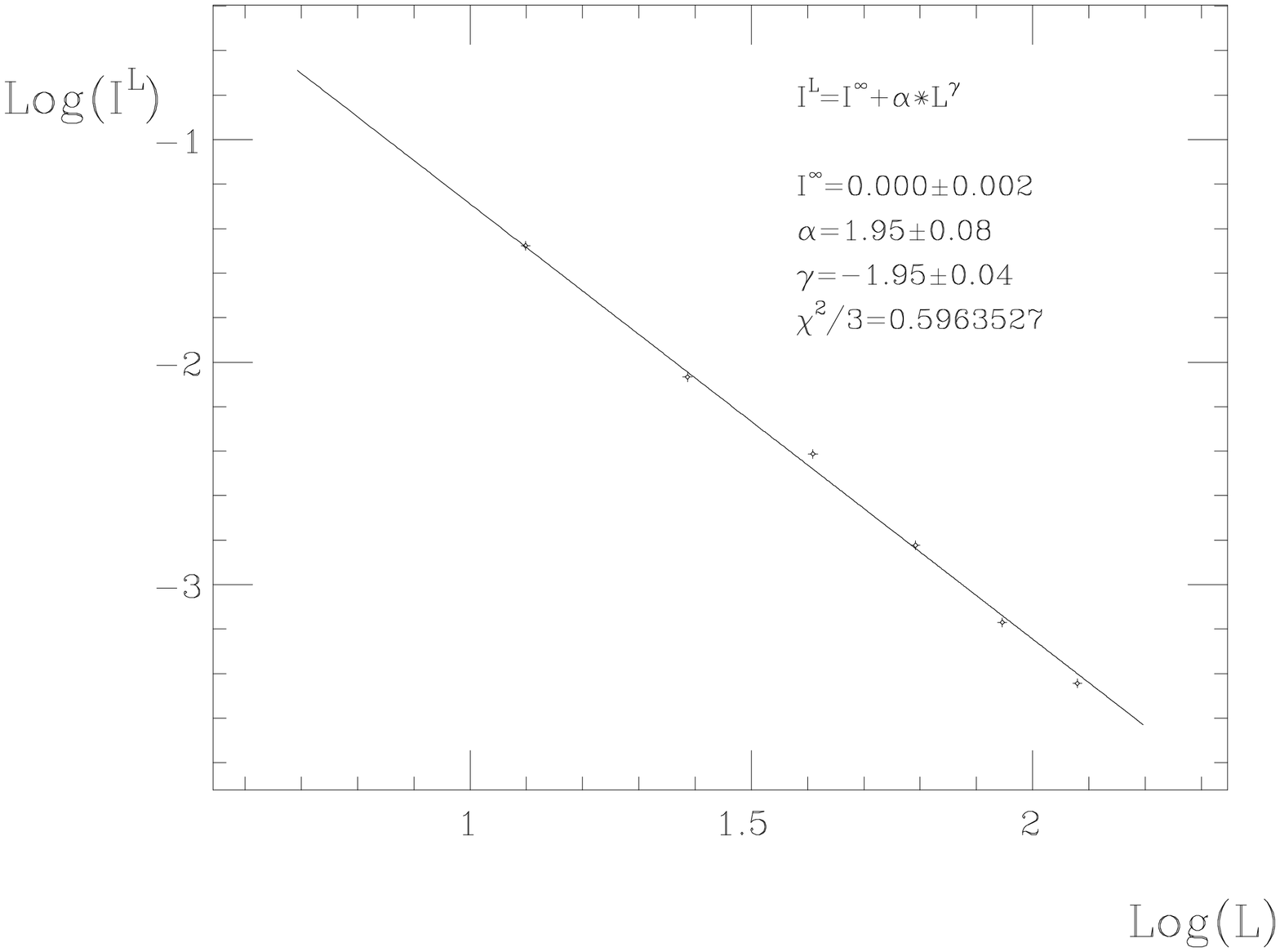
,bbllx=50pt,bblly=98pt,bburx=545pt,
bbury=600pt,width=280pt,height=240pt,angle=90}}

\vspace{-.5cm}
\caption[986]{Evolution of the integral
$I_L=\int_{-1}^{1} P_L(q_L)(q_L-0.10)^{2}dq_L$ with the lattice size
 in log-log scale}
\label{f3}
\end{figure}

\begin{figure}[h]
\centerline{
\epsfig{figure=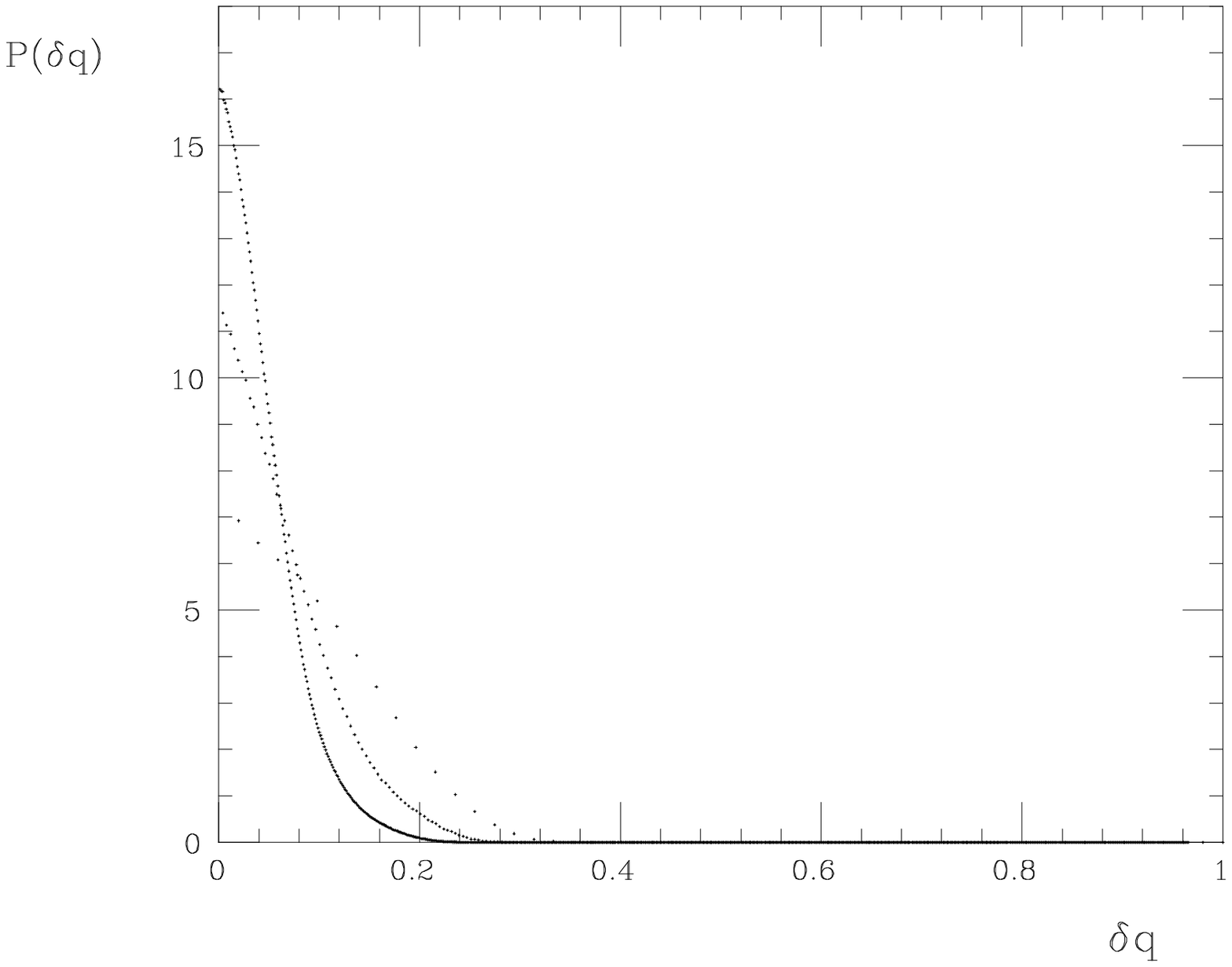
,bbllx=50pt,bblly=98pt,bburx=545pt,
bbury=600pt,width=320pt,height=320pt,angle=90}}
\caption[986]{Ultrametric feature of the EA model for lattice 
of side L=4, 6 ,8. We plot the probability of the
 difference between the middle and the 
smaller overlap $\delta q=q_{mid}-q_{min}$ when the largest fall inside
the range $[2/5q_{EA},q_{EA}]$ }
\label{o1}
\end{figure}

\end{document}